%% file: main.tex
\documentclass[conference,compsoc]{IEEEtran}
\ifCLASSOPTIONcompsoc
  \usepackage[nocompress]{cite}
\else
  \usepackage{cite}
\fi
\ifCLASSINFOpdf
\else
\fi
\usepackage{amsmath}
\usepackage[skip=1pt]{caption}
\usepackage[labelformat=simple, skip=1pt]{subcaption}
\usepackage{multirow}
\usepackage{tabularx}
\usepackage{xcolor}
\usepackage{array,longtable}
\usepackage{booktabs}
\usepackage{float}
\usepackage{graphicx}
\usepackage{makecell}
\usepackage{paralist}
\usepackage{adjustbox}
\usepackage{svg}
\usepackage{csquotes}
\usepackage{pifont}
\usepackage{cancel}
\usepackage{soul} 
\usepackage{hyperref}

\usepackage{wasysym}
\usepackage{enumitem}
\usepackage{color}
\usepackage{forloop}
\usepackage{ifthen}

\newcounter{row}
\makeatletter
\@addtoreset{subfigure}{row}
\makeatother

\newcommand{\cmark}{\textcolor{green!80!black}{\ding{52}}}

\newcommand{\cmarkWrong}{\textcolor{orange}{\bcancel{\ding{52}}}}
\newcommand{\xmark}{\textcolor{red}{\ding{54}}}

\newcommand{\Qq}[1]{\textbf{#1}}

\newcommand{\QOC}{$\ocircle$}

\newcounter{qr}

\newcommand{\QratingC}[1]{\QOC\forloop{qr}{1}{\value{qr} < #1}{---\QOC}}

\newcommand{\Qline}[1]{\noindent\rule{#1}{0.6pt}}

\newcounter{ql}

\newenvironment{Qlist}{%

\begin{itemize}[leftmargin=1.5em,topsep=-.5em]
}{%
\end{itemize}
}

\newenvironment{QlistC}{%

\begin{itemize}[leftmargin=1.5em,topsep=-.5em]
}{%
\end{itemize}
}

\newlength{\qt}

\newcounter{itemnummer}
\newcommand{\Qitem}[2][]{
\ifthenelse{\equal{#1}{}}{\stepcounter{itemnummer}}{}
\ifthenelse{\equal{#1}{a}}{\stepcounter{itemnummer}}{}
\begin{enumerate}[topsep=2pt,leftmargin=2.8em]
\item[\textbf{\arabic{itemnummer}#1.}] #2
\end{enumerate}
}

\definecolor{bgodd}{rgb}{0.8,0.8,0.8}
\definecolor{bgeven}{rgb}{0.9,0.9,0.9}
\newcounter{itemoddeven}
\newlength{\gb}
\newcommand{\QItem}[2][]{
\setlength{\gb}{\linewidth}
\addtolength{\gb}{-5.25pt}
\ifthenelse{\equal{\value{itemoddeven}}{0}}{%
\noindent\colorbox{bgeven}{\hskip-3pt\begin{minipage}{\gb}\Qitem[#1]{#2}\end{minipage}}%
\stepcounter{itemoddeven}%
}{%
\noindent\colorbox{bgodd}{\hskip-3pt\begin{minipage}{\gb}\Qitem[#1]{#2}\end{minipage}}%
\setcounter{itemoddeven}{0}%
}
}

\newcommand{\parheading}[1]{\noindent{}\textbf{{#1}}}
\newcommand{\eg}{e.g.,}
\newcommand{\ie}{i.e.,}

\newcommand{\etal}{et al.}
\newcommand{\wrt}{w.r.t.}
\newcommand{\tool}{\textsc{VBIT}}

\newif\ifanswers
\answerstrue 

\newif\ifPETSanswers

\newif\ifFactors

\begin{document}
%
\title{\tool{}: Towards Enhancing Privacy Control Over IoT Devices}

\author{\IEEEauthorblockN{Jad Al Aaraj\IEEEauthorrefmark{1},
Olivia Figueira, Tu Le, Isabela Figueira, Rahmadi Trimananda, and
Athina Markopoulou}
\IEEEauthorblockA{University of California, Irvine\\
Irvine, CA\\
Email: \IEEEauthorrefmark{1}jalaaraj@uci.edu}}


%


\maketitle

\input{sections/00-abstract}


%
\IEEEpeerreviewmaketitle

\input{sections/01-introduction}

\input{sections/02-relatedwork}
\input{sections/03-system_design}
\input{sections/04-study_methods}
\input{sections/05-results}
\input{sections/06-discussion}

\ifCLASSOPTIONcompsoc
  \section*{Acknowledgments}
\else
  \section*{Acknowledgment}
\fi

This research is supported by the National Science Foundation (NSF) under awards CNS-1815666 (``SaTC: CORE: Small: Collaborative: A Multi-Layer Learning Approach to Mobile Traffic Filtering") and CNS-1956393 (``SaTC: Frontiers: Collaborative: Protecting Personal Data Flow on the Internet").



%


\bibliographystyle{IEEEtran}
\bibliography{main}

\appendices
\input{sections/appendix}

\end{document}

%% file: sections/00-abstract.tex
\begin{abstract}
Internet-of-Things (IoT) devices are increasingly deployed at home, at work, and in other shared and public spaces. IoT devices collect and share data with service providers and third parties, which poses privacy concerns.
Although privacy-enhancing tools are quite advanced in other application domains (\eg~ advertising and tracker blockers for browsers), users currently have no convenient way to know or manage what and how data is collected and shared by IoT devices.
In this paper, we present \tool{}, an interactive system combining Mixed Reality (MR) and web-based applications that allows users to: (1) uncover and visualize tracking services by IoT devices in an instrumented space and (2) take action to stop or limit that tracking. 
We design and implement \tool{} to operate at the network traffic level, and we show that it has negligible performance overhead and offers both flexibility and good usability.  
We perform a mixed-method user study consisting of an online survey and an in-person interview study.
We show that \tool{} 
users appreciate \tool{}'s transparency, control, and customization features, and they become significantly more willing to install an IoT advertising and tracking blocker after using \tool{}.
In the process, we obtain design insights that can be used to further iterate and improve the design of \tool{} and other systems for IoT transparency and control.
\end{abstract}

%% file: sections/01-introduction.tex
\section{Introduction}
The rise of Internet of Things (IoT) devices has introduced an abundance of smart devices in instrumented spaces at home, work, and in public. 
The number and variety of IoT devices is continuously increasing,
and the types of data collected are becoming increasingly sensitive, from personal identifiers and user activity data to ambient sensor measurements, voice, and video. 
Such data can be bought, sold, or used without users' knowledge, enabling tracking of users' daily activities and risking security vulnerabilities if malicious actors gain access. 
 
While mature ecosystems such as web~\cite{ublock, adblock} and mobile~\cite{antmonitor, adguard_android} have well-developed anti-tracking technologies~\cite{filter_lists, autofr}, the IoT domain currently lags behind \wrt{} transparency and control of personal data. Today, people walking into spaces instrumented with IoT devices lack convenient means to both understand and control what data is collected and by which entity. IoT devices routinely collect both sensor and personal data and share it with service providers, automation systems, advertisers, and trackers~\cite{iot_traffic_wild}. Yet these data collection practices  are not clearly conveyed to users, neither by the IoT user interfaces, nor by their privacy policies~\cite{litman_navarro_policies_2019}. Some of these  challenges are common with other application domains, but are further exacerbated for IoT devices by limitations of screen sizes, modes of interaction, and sensitivity of data collected.

A universal approach is to use the edge router as a vantage point to observe network traffic and data that is both collected by local devices and sent outside the local network. The popular Pi-hole is widely used to block DNS requests, thus blocking entire domains contacted by any connected devices in the local network. There are some blocklists such as~\cite{Firebog-SmartTV, Firebog-AmazonFireTV} that are curated for IoT that could be deployed at the edge of the network at the domain level. On the research side, both Privacy Plumber~\cite{ndss_2023_ar_privacy} and IoT Inspector~\cite{iot_inspector_2020} also analyze and block network traffic.

Overall, existing IoT privacy solutions are currently limited:  they offer coarse granularity, limited  functionality, and are complicated for non-expert users.

In this paper, we present a system, \textbf{\tool{}} (\textbf{V}isualizing and \textbf{B}locking of \textbf{I}oT \textbf{T}rackers), that allows the owner of an instrumented space (at home, work, or public spaces) to provide transparency and control to people in that space by revealing the IoT devices present and providing an interactive interface for visualization and control of tracking. The IoT devices can range from simple devices (\eg~ smart plugs or smart bulbs) to more complicated ones (\eg{} voice assistants and smart TVs), as long as they are connected to the Internet through a router.
We rely on network traffic analysis and blocking to provide data transparency and privacy control through visualization and interaction.

\parheading{The \tool{} System.}
\tool{}
consists of the components depicted in Figure~\ref{fig:system_overview}.
First, it monitors the outgoing network traffic generated by IoT devices, reports the entities that collect data from the connected devices, determines the organizations to which the domains belong, and determines whether the domains are associated with ads and tracking services (ATS), which we refer to as trackers.
Second, \tool{} provides both an {\bf interactive Mixed Reality (MR)} and a {\bf Web-based app} for visualizing and blocking that IoT traffic. The interactive mobile app with MR visualization overlays the IoT device with a virtual panel and enables users to quickly inspect and block IoT device trackers.
The web-based app generates and displays figures and graphs to visualize summaries and statistics of the devices' traffic in real-time in much finer granularity than previously possible~\cite{iot_inspector_2020}; 
it enables in-depth analysis and caters to a broad range of user preferences, ensuring comprehensive insights.

The prototype implementation of \tool{} leverages Pi-hole that runs on a Raspberry Pi (RPi) device. However, \tool{} can be implemented on any linux-based router, such as access points running OpenWRT~\cite{openwrt} or FreeBSD~\cite{freebsd}.
Our performance evaluation of the RPi prototype shows that \tool{} is more effective than prior work~\cite{ndss_2023_ar_privacy,iot_inspector_2020} as it incurs negligible overhead on the network (compared to Privacy Plumber~\cite{ndss_2023_ar_privacy}) with acceptable power consumption. It also provides finer granularity by allowing blocking per domain. We plan to release \tool{} as a plug-and-play tool, envisioning that it can be used by end-users (\eg ~ people at their home, Pi-hole~\cite{Pihole} users),
administrators (\eg~ space  and building administrators),
experts (\eg~IoT blocklist authors), and researchers (\eg~ to improve blocklist creation~\cite{autofr} specifically for IoT~\cite{iot_trim_list}).

\parheading{User Study.}
We also conduct a mixed-method user study
comprising of an online survey and an in-person interview study with 200 and 18 participants, respectively. The goal was to evaluate \tool{}'s usability and effectiveness in empowering and educating users about IoT privacy, as well as to learn from users' feedback for future design iterations. 
Key findings corresponding to the user study questions (\textbf{UQ0-3}) in Section~\ref{sec:results}, include the following: 
\begin{itemize}
    \item[\textbf{UQ0:}]{\em Before} learning about \tool{}\footnote{UQ0 and UQ1-3 are aimed at understanding our participants' perceptions and expectations about IoT privacy {\em before} and {\em after} learning about \tool{}, respectively. UQ0 has been explored in prior work, agreeing with prior observations, as discussed in Section~\ref{sec:relatedwork}. We still include UQ0 in our study design to (i) confirm that our participants represent a diverse range of privacy perceptions in line with prior work and (ii) to compare our participants' attitudes before and after learning about \tool{}.}, users generally desire greater control and transparency regarding IoT privacy but have varying levels of awareness and concern.
    
    Factors that affect users' attitudes include the kinds of IoT devices users own, and the effect blockers have on the usability of IoT devices.

    \item[\textbf{UQ1:}] {\em After} learning about and using \tool{}, users become more aware of IoT tracking and are more willing ($p=0.002$) to install an IoT ad/tracker blocker.

    \item[\textbf{UQ2:}]  {\em Usability.} \tool{} currently achieves ratings up to 82.5/100 on the system usability scale (SUS), which is considered good/excellent~\cite{sus_bangor}. Users particularly appreciated \tool{}'s granular control over their IoT privacy through custom tracker blocking.
    
    \item[\textbf{UQ3:}] \emph{Design Insights.} In the process, we obtained insights to further evolve and improve \tool's{} design w.r.t. transparency, control, and customization of  IoT ad/tracker blockers. For example, users prefer to use the MR app for quick actions and the web app for more in-depth analysis, and users want notifications when a new tracker is contacted.
    
\end{itemize}

Our contributions lie both in the system design and implementation and in the lessons learned from the user study. We believe this is a first but much needed step in advancing data transparency and privacy control in IoT-instrumented spaces. The structure of the rest of the paper is as follows: Section~\ref{sec:relatedwork} reviews related work and describes \tool{}'s contributions in this space.
Section~\ref{sec:design} presents the design of \tool{}'s components and performance evaluation.
Section~\ref{sec:user-study} presents the design of our user study, including both survey and interview studies, and Section~\ref{sec:results} presents the results of the user study. 
Section~\ref{sec:discussion} provides discussion and concludes the paper.

%% file: sections/02-relatedwork.tex
\section{Background and Related Work}\label{sec:relatedwork}

In this section, we review studies of user perspectives on IoT privacy (Section \ref{sub_sec_rel:user_perspective}), systems for visualization of IoT devices  (Section \ref{sub_sec_rel:iot_viz}), and systems specifically for enhancing IoT privacy by enabling control (Section \ref{sub_sec_rel:iot_sys}), and we discuss how prior findings motivate our design. We also summarize \tool{}'s contributions and differences from prior work (Section \ref{sec:perspective}).

\subsection{\textbf{User Perspectives on IoT Privacy}}\label{sub_sec_rel:user_perspective}

Prior studies have assessed users' attitudes and expectations regarding IoT privacy from several perspectives. With regards to users' acceptance criteria and motivations for adopting IoT devices as well as privacy protection mechanisms, prior work has found that users are driven by several factors, including convenience (\eg{} automation for everyday tasks), utility (\eg{} health and safety monitoring), cost, and IoT devices' privacy and security risks~\cite{internet_of_what_xinru_2018, coughlan_exploring_2012, IoT_norms_apthorpe_2018, emami_iotpurchase_2023, spaf_das_2023}. Several studies have also explored users' concerns regarding IoT privacy and security and the actions they take to protect themselves, but users' concerns are often influenced by their level of technical knowledge, awareness of privacy and security risks, and desire for IoT utility and convenience, all of which inform users' personal privacy-utility trade-off decisions~\cite{user_iot_concerns_zeng_2017, smart_home_privacy_perceptions_zheng_2018, privacy_iot_world_emani_naeini_2017, voice_assistant_privacy_dunbar_2021}. IoT privacy concerns are further exacerbated in multi-user smart spaces, such as visitors of smart homes and commercial smart spaces. In multi-user smart spaces, in addition to concerns that are similar to general IoT privacy concerns, visitors are often unable to take actions to protect their privacy~\cite{zeng_roesner_2019, le2023smartbuilding, privacy_tea_marky_2022, smarthome_visitors_marky_2020, smart_speaker_sharing_meng_2021}. Additionally, Jakobi \etal{}~\cite{longitudinal_iot_privacy_jakobi_2018} conducted a longitudinal study with users across 12 smart homes to investigate users' strategies for IoT control and monitoring, and they found that users initially used the provided web-based dashboard that had more in-depth information, but over time, they tended to access the interface from their mobile device when needed, thus demonstrating the need for a multi-modal interface, which motivates the design of \tool{}. We observe that the user privacy concerns from our participants are in line with prior work, and we introduce the participants' attitudes regarding IoT privacy before and after learning about an IoT blocker, represented in our work as \tool{}, in Section~\ref{sec:results}. 

\begin{table*}[t]
    \centering
    \footnotesize
    \captionsetup{font={small}}
    \caption{\tool{} compared to prior IoT privacy methods. The orange check mark with a slash (\cmarkWrong) indicates a partial solution.}
    \begin{tabular}
    {>{\centering\arraybackslash}p{.1\linewidth}|>{\centering\arraybackslash}p{.1\linewidth}>{\centering\arraybackslash}p{.08\linewidth}>{\centering\arraybackslash}p{.05\linewidth}|>{\centering\arraybackslash}p{.315\linewidth}|>{\centering\arraybackslash}p{.05\linewidth}>{\centering\arraybackslash}p{.075\linewidth}>{\centering\arraybackslash}p{.05\linewidth}}
    \specialrule{.1em}{0em}{0em}
    \textbf{System} & \textbf{Multimodal Interface} & \textbf{Tracking\newline Visualization} & \textbf{Privacy\newline Control} & \textbf{System Performance Evaluation}  & \textbf{User Study} & \textbf{Interview Study Size} & \textbf{Survey  Size} \\
    \specialrule{.1em}{0em}{0em}
    IoT Privacy Labels\cite{iot_label_2020} & \xmark(Paper) & \xmark & \xmark & (N/A)  no system implementation  & \cmark & 22 & - \\
    \specialrule{.05em}{0em}{0em}
    PriKey\cite{prikey}  & \xmark (Physical Handheld Object) & \xmark & \cmarkWrong &  (N/A) no system implementation & \cmark & 16 & - \\
    \specialrule{.05em}{0em}{0em}
    PriView\cite{priview} & \xmark(AR) & \cmark & \xmark & Device detection with an average loss of 0.4143. & \cmark & 24 & - \\
    \specialrule{.05em}{0em}{0em}
    IoT Inspector\cite{iot_inspector_2020} & \xmark (Web) & \cmark & \xmark & -(Not reported) & \xmark & - & - \\
    \specialrule{.05em}{0em}{0em}
    Privacy Plumber\cite{ndss_2023_ar_privacy} & \xmark (AR) & \cmark & \cmarkWrong & 85\% network bandwidth reduction & \cmark & 6 & - \\
    \specialrule{.1em}{0em}{0em}
    \textbf{\tool{}} & \cmark (MR + Web) & \cmark & \cmark & 3\% network bandwidth reduction & \cmark & 18 & 200 \\
    \specialrule{.1em}{0em}{0em}
    \end{tabular}
    \label{tab:related_works_comp}
\end{table*}

\subsection{\textbf{IoT Visualization}}\label{sub_sec_rel:iot_viz}
The research community has worked to create IoT visualization systems with extended reality (XR) technology, which includes augmented, virtual, and mixed reality (A/V/MR), and web-based apps with the goal of improving user awareness of both IoT device functionalities and potential privacy risks.
For example, prior work have developed several systems that use XR technologies to visualize IoT device information: Lumos~\cite{sharma_lumos_2022} is an AR system that enables a user to walk around a room and visualize information about hidden IoT devices. VR Binoculars~\cite{vr_binoculars} is an MR application for VR headsets that overlays interactive visualizations of IoT device data, such as sensor readings and notifications, on the devices in the environment. Fuentes \etal{}~\cite{SAR.IoT} proposed an AR tool for mobile devices to display real-time IoT sensor information, and Clark \etal{}~\cite{articulate} proposed an AR system called ARticulate that allows users to determine IoT device functionalities and capabilities in unfamiliar spaces. Regarding web-based systems, Jeong \etal{}~\cite{jeong_joo_hong_shin_lee_2015} proposed a system called AVIoT that allows users to create 3D models of their smart spaces and define their IoT devices' behaviors, such as triggering devices based on sensor readings, through abstracted visual programming.
The mentioned prior works are largely focused on visualizing IoT devices, functionality, and data to increase user awareness, with some also providing interaction mechanisms for utility purposes. While the prior works help to inform the design of the visualization components in \tool{}, we focus specifically on privacy visualization and control.
This is known to be key for user understanding and engagement~\cite{unwin_vis_2020}  with  privacy-enhancing technologies.

\subsection{\textbf{IoT Privacy and Control}}\label{sub_sec_rel:iot_sys}
Researchers have proposed systems and mechanisms to enhance privacy and control for IoT devices through various means.
We summarize the most relevant ones in Table~\ref{tab:related_works_comp} and compare them to \tool{}.
Table~\ref{tab:related_works_comp} describes whether each work includes a multi-modal interface (\eg{} MR, web, other), IoT tracking visualization, IoT privacy controls (\eg{} whether features are available to enable users to actively restrict data collection), a system performance evaluation, and a user study (also indicating whether a mixed-method study was conducted).
Emami-Naeini \etal{}~\cite{iot_label_2020} proposed {\em privacy and security labels} for IoT devices to provide users an opportunity for informed decision-making. However, once users decide to purchase the IoT device, they still lack control mechanisms to protect themselves from IoT tracking.
Delgado Rodriguez \etal{} proposed {\em PriKey}~\cite{prikey}, a wizard-of-oz prototype that allows control of audio, video, and location data through a physical key, and conducted a user study to understand users' perceptions regarding the prototype. However, PriKey did not include a deployable implementation.
Prange \etal{}~\cite{priview} developed {\em PriView}, which detects IoT devices using a thermal camera through an AR interface, and the authors proposed five different visualization options to alert the users of the presence of possibly invasive IoT devices. The limitations of PriView include a relatively low detection rate of devices and a lack of controls to stop IoT devices' data collection. 

The closest work to \tool's{} web app and network-based approach is {\em IoT Inspector}~\cite{iot_inspector_2020}, by Huang \etal{}, which intercepts network traffic from IoT devices using ARP spoofing and reports aggregate measurements over time through a web-based app. However, ARP spoofing has two key weaknesses: (1) it increases network traffic and congestion because it floods the network with false ARP messages and (2) apps that use ARP spoofing are not allowed to be published on app stores due to prohibitions on malicious activities and network disruption. Furthermore, the IoT Inspector web app displays information regarding which domains the IoT devices are contacting but provides no method to block any of the domains. 
The closest work to \tool{}'s MR app is {\em Privacy Plumber}, a workshop paper by Cruz \etal{}~\cite{ndss_2023_ar_privacy}, which also uses the camera on a mobile device to scan a marker and display a visualization through an AR interface. The drawbacks of Privacy Plumber are the single mode of visibility (\ie{} AR app only), the very high bandwidth reduction (85\%), the lack of selective privacy controls (\eg{} blocking of domains blocks internet connection entirely), and their user study size (\ie{} only 6 participants and a few questions).

Another popular network traffic-based system is {\em Pi-hole}~\cite{Pihole}, which blocks advertising domains on a network-wide level. While Pi-hole provides some statistics (\eg{} total queries and queries blocked), its primary focus is on blocking unwanted content rather than analyzing traffic on a per-device basis. Furthermore, Pi-hole neither labels domains as trackers/non-trackers nor provides a straightforward method to block such domains. Users have to access the Pi-hole dashboard and add filter lists that include a large number of domains to block, leading to possible breakage of their IoT devices without a way to see what specific filter list rules resulted in breakage.
Due to their similar blocking features, \tool{} utilizes Pi-hole to enable users to block selected domains. \tool{} enhances Pi-hole by providing better visualization, detailed statistics, and domain classification on a per-device basis. Additionally, \tool{} can be implemented as an extension on top of Pi-hole, providing easy integration for users who already have Pi-hole installed.

\subsection{Our Work in Perspective \label{sec:perspective} }
While traditional ad blockers are commonly available for web browsers, there is a noticeable gap in the availability of easy-to-use IoT ad blockers. The current available solutions either have bad performance or require some technical knowledge to set up and manage, which can be a barrier for average users. Building on the findings of the aforementioned prior work and shortcomings of previous IoT privacy-enhancing systems, we designed and developed a system for IoT traffic visualization and tracker blocking.
To the best of our knowledge, \tool{} combines, for the first time, several desired characteristics:

(1) \tool{} is based on network traffic, thus leveraging a universal vantage point for visibility and blocking of trackers. It has negligible network overhead (compared to 85\% overhead in Privacy Plumber~\cite{ndss_2023_ar_privacy}, which uses ARP spoofing) and the capability for selective blocking of domains while maintaining device functionality and internet connectivity (compared to blocking all internet connection, such as in Privacy Plumber~\cite{ndss_2023_ar_privacy}).

(2) \tool{} is a multi-modal system with both MR and web apps, enabling users to benefit from both kinds of interaction mechanisms: the MR app allows for quick exploration and the web app allows for a deeper dive into the details of each device (as opposed to aggregate statistics, such as in IoT Inspector~\cite{iot_inspector_2020} and Pi-hole~\cite{Pihole}).

(3) \tool{} provides both privacy control through blocking and informative visualizations, and it is a functional prototype (as opposed to a design or concept, such as in \cite{prikey}).

(4) \tool{} is evaluated through a mixed-method user study (in-person interviews and an online survey) to assess its usability with feedback from users with diverse technical backgrounds.

(5) Compared to the basic functionality of Pi-hole, \tool{} offers user-friendly interactive interfaces for visualization and blocking, per-device analysis, and labeling of candidate tracker domains not necessarily for blocking but to inform users' decisions about what domains to block.
\tool{} improves upon prior work, providing users with a multi-modal MR and web-based system for IoT privacy transparency and control.

(6) Last but not least, \tool{} provides a finer-grained granularity of blocking (per domain), which is an improvement over the traditional bulk blocking practice. We envision that \tool{} can be used by end-users (\eg~ people at their home, Pi-hole users), administrators (\eg~ space  and building administrators), experts (\eg~IoT blocklist authors), and researchers (\eg~ to improve blocklist creation~\cite{autofr} specifically for IoT~\cite{iot_trim_list}). This is a much needed first step towards advancing IoT specific blocklists.

%% file: sections/03-system_design.tex
\section{\tool{} Design and Implementation}\label{sec:design}

In this section, we present the \tool{} system, whose overview is also depicted in Figure~\ref{fig:system_overview}. We envision that \tool{} will be deployed by the owner/administrator of a smart space to provide transparency and control over the IoT devices in that space.  We rely on a universal vantage point, the edge router, to monitor and block the outgoing network traffic from IoT devices in that space. The user interacts with VBIT through two apps: one mobile MR and one web app, in order to visualize and selectively block traffic to tracker domains. In this setting, all VBIT components trust each other and work together: the administrator of the smart space owns the devices and the router, which talk to the MR and web apps through a local server, to offer the user a comprehensive visualization and options for blocking IoT tracking.

\ifanswers
\begin{figure}[!t]
\centering
\includegraphics[width=\linewidth] {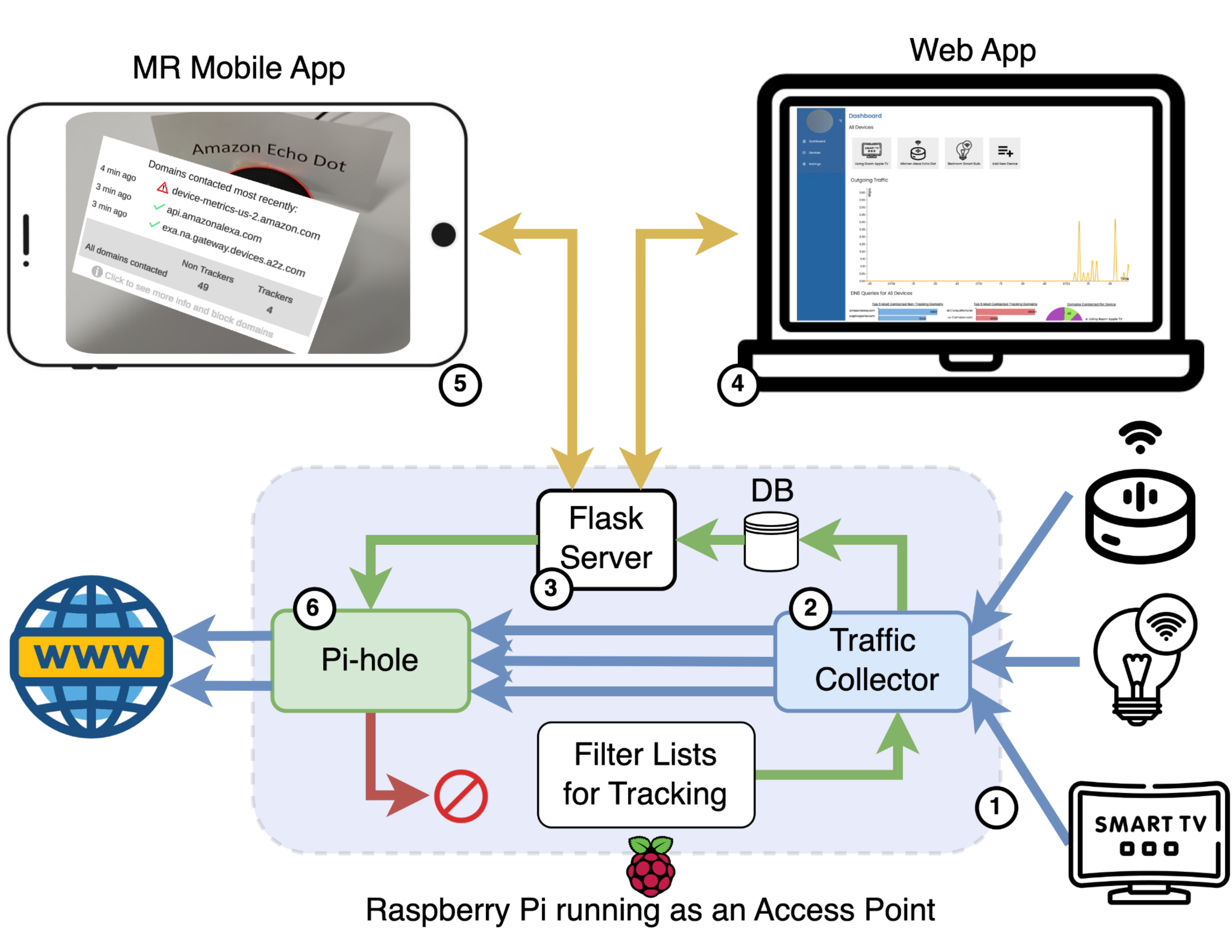}
\captionsetup{font={footnotesize}}
\caption{\tool{} Overview. \tool{} consists of the following components: 
\textbf{(1)} \textit{Raspberry Pi (RPi)} running as an Access Point (AP) for the IoT devices to connect to; on the RPi,
\textbf{(2)} \textit{traffic collector} collects all network traffic from the IoT devices, records the traffic and its metadata in the \textit{database}, labels each destination as tracker or non-tracker based on \textit{filter lists for tracking}, and forwards the traffic;
\textbf{(3)} \textit{Flask server} communicates (\ie{} via HTTP requests) with \tool{}'s 
\textbf{(4)} \textit{MR mobile} and \textbf{(5)} \textit{web apps}, and retrieves information from the database to be displayed on the apps (\ie{} destinations contacted in the network traffic and whether each is a tracker or non-tracker), which will also send traffic blocking decisions based on user's interaction with the app;
\textbf{(6)} \textit{Pi-hole} will block certain traffic for tracking destinations (\eg{} {\color{red} red colored} arrow) depending on the user's decision when using the apps.}
\label{fig:system_overview}
\end{figure}
\fi

\subsection{Network Traffic Interception and Blocking}

\subsubsection{\textbf{Network Interception}}
VBIT can be installed on top of any linux-based router (such as devices with OpenWrt). In this paper, we develop a prototype based on a Raspberry Pi (RPi), configured as a wireless access point (AP). The IoT devices on the network are then able to connect to the RPi which allows us to intercept all the thru-traffic for further analysis. This is achieved by developing our own traffic collector module that collects the network traffic from the devices and stores the metadata in a database. Our traffic collector module uses Scapy~\cite{scapy} to capture the network traffic in real-time. To capture DNS packets, we filter on port 53 and check the DNS ANCount. The DNS ANCount indicates the number of resource records found in the answer section of a DNS response. By examining the answer section, we can determine the domain contacted by the device, log the domain information in the database, update the timestamp accordingly, and map IP addresses to domains. The traffic collector also logs the real-time connections by capturing the TCP or UDP packets generated by the IoT device. We then use the destination IP addresses of those connections to get the domain name contacted through the IP address-to-domain mappings. We rely on the TCP/UDP connections to detect the real-time domains contacted instead of the latest DNS query since the device might have used the local DNS resolver cache to obtain the IP address of the internet resource.

\subsubsection{\textbf{Filter Lists}}
In order to label a domain as a potential tracker or not, we rely on filter lists.
Since no single general IoT blocklist exists, we compiled several well-known filter lists used by adblockers on the web and on certain types of smart IoT devices\footnote{
Designing the best IoT filter list is out of scope for this work. Instead, we use state-of-the-art lists as a proof of concept for our prototype.  We combine the ``Tracking \& Telemetry Lists'' from firebog \cite{Firebog}.
The lists are 
Easyprivacy \cite{Firebog-Easyprivacy}, 
Prigent-Ads \cite{Firebog-Prigent-Ads},
Extra rules for StevenBlack's hosts project filter list \cite{Firebog-FadeMind},
crazy-max WindowsSpyBlocker \cite{Firebog-WindowsSpyBlocker},
frogeye firstparty-trackers-hosts \cite{Firebog-firstparty-trackers-hosts},
ads-and-tracking-extended \cite{Firebog-ads-and-tracking-extended},
PiHoleBlocklist android-tracking \cite{Firebog-android-tracking},
PiHoleBlocklist SmartTV \cite{Firebog-SmartTV},
PiHoleBlocklist AmazonFireTV \cite{Firebog-AmazonFireTV},
and notrack-blocklist \cite{Firebog-notrack-blocklist}.}.
We use these filter lists to label the contacted domains as potential trackers, we store that information in the database and we present it to the user, in the MR and web apps. The user can then decide whether to block a domain or not.

\subsubsection{\textbf{Back-end Server}} We develop a Flask server~\cite{Flask} to interact with the MR and web apps. This Flask server communicates, sends information stored in the database to the apps for display purposes, and accepts requests to block/unblock certain tracker domains.

\subsubsection{\textbf{Network Blocking}}  We install Pi-hole~\cite{Pihole} on the RPi, an application that acts as a DNS sinkhole, allowing us to control the outgoing DNS requests and block any of the contacted domains. We also used Pi-hole in \tool{} as the DHCP server that assigns IP addresses to the connected devices. We decided to use Pi-hole since it allows blocklisting domains dynamically. The network traffic passing through the RPi passes into the Pi-hole module where the DNS query of any blocklisted domain can blocked. 

\ifanswers
\begin{figure*}[h]
    \renewcommand{\thesubfigure}{(\alph{row})(\roman{subfigure})}
    \centering 
    \setcounter{row}{1}
    \begin{subfigure}{0.32\textwidth}
        \includegraphics[width=\linewidth]{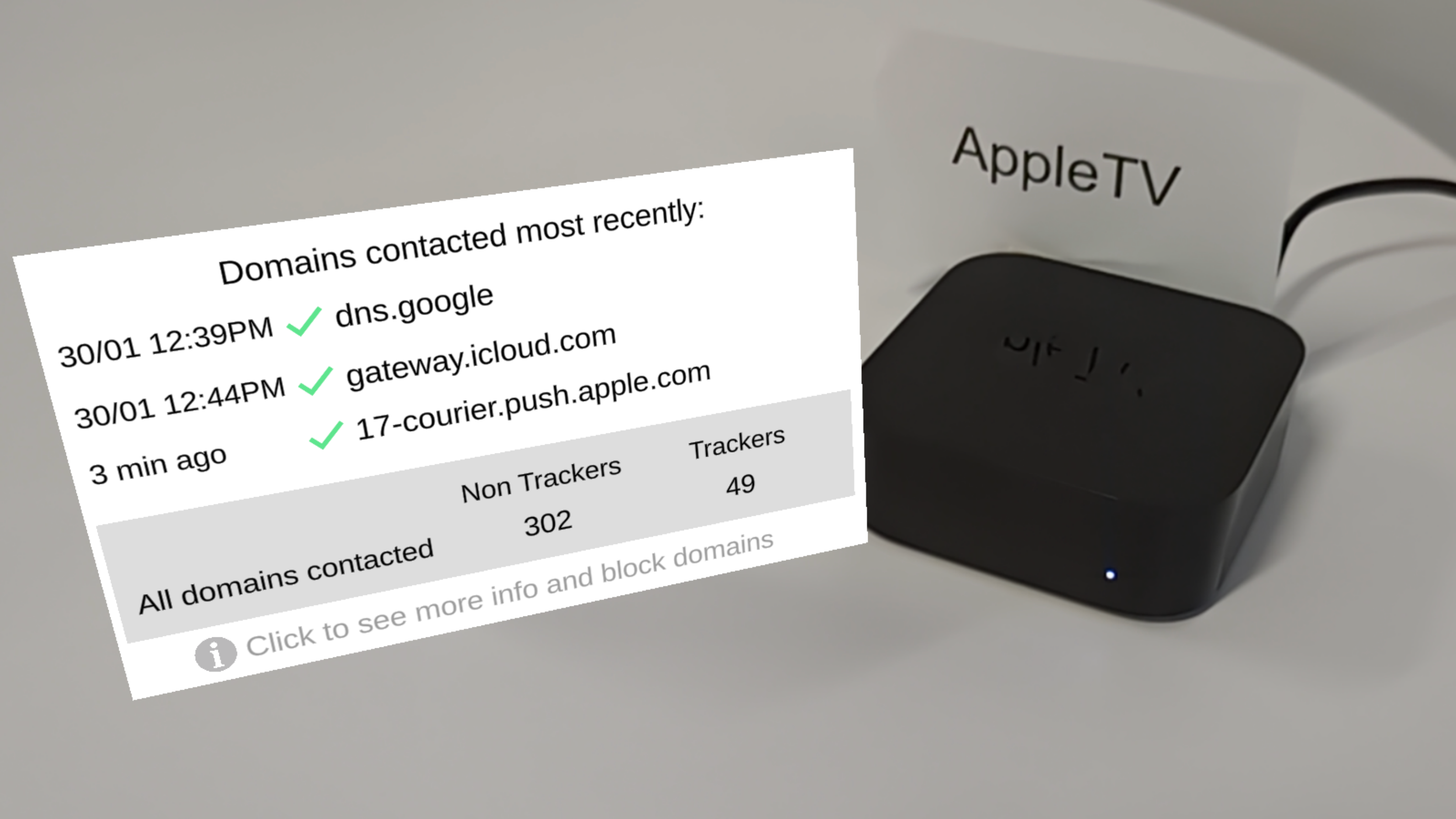}
        \caption{Apple TV Recent Information}
        \label{fig:3x3_apple_1}
    \end{subfigure}
    \hfill
    \begin{subfigure}{0.32\textwidth}
        \includegraphics[width=\linewidth]{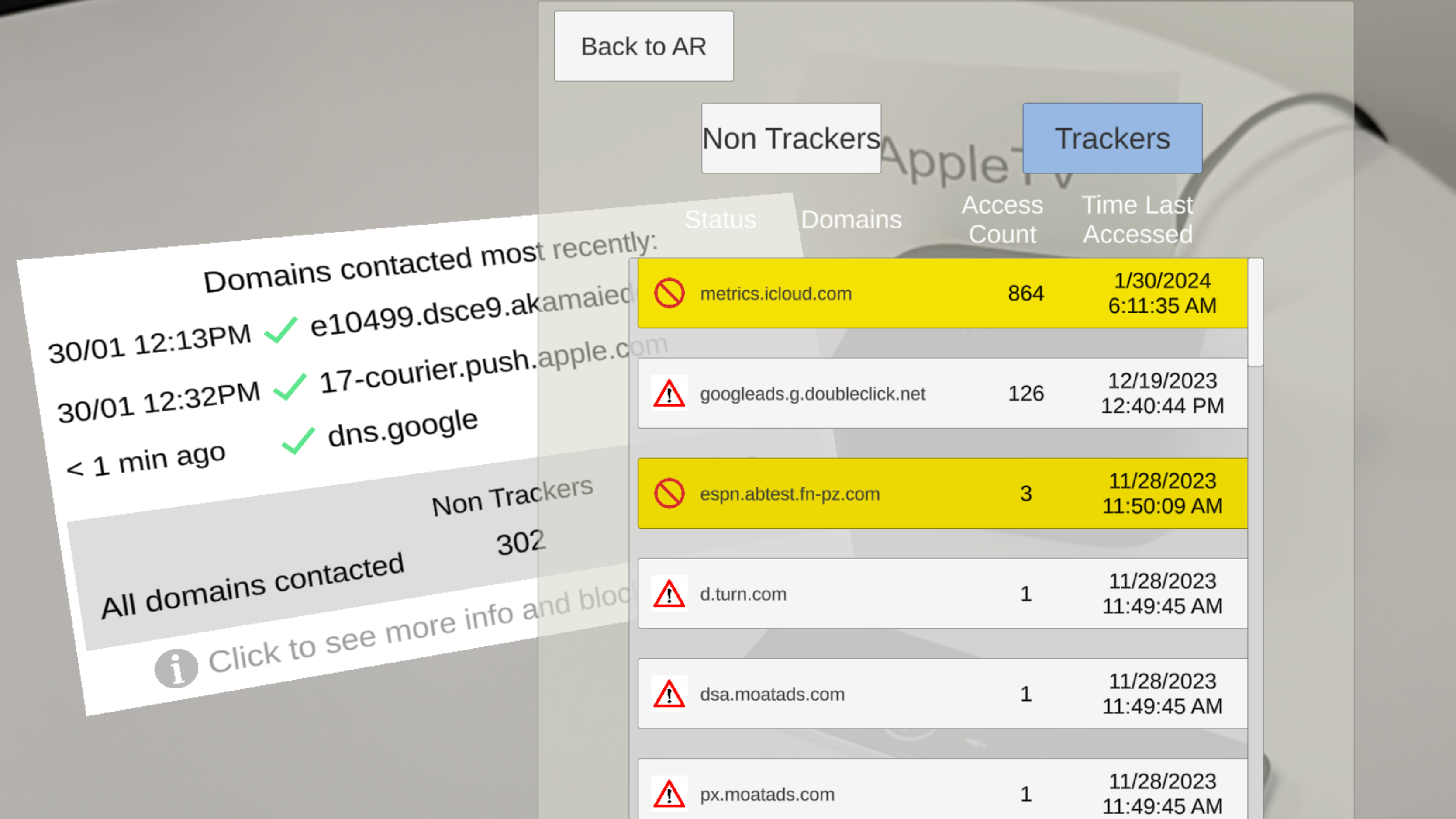}
        \caption{Apple TV Trackers}
        \label{fig:3x3_apple_2}
    \end{subfigure}
    \hfill
    \begin{subfigure}{0.32\textwidth}
        \includegraphics[width=\linewidth]{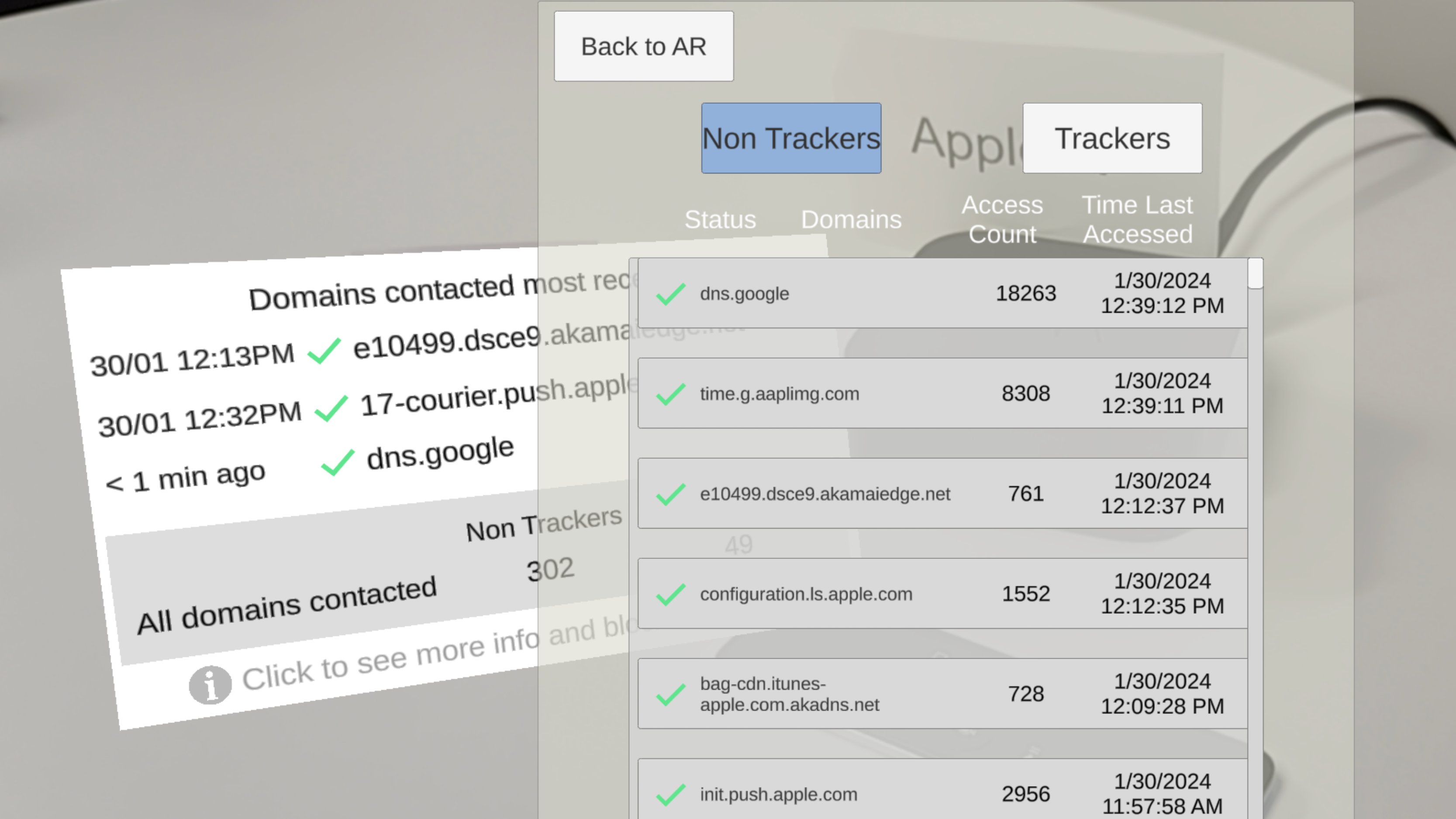}
        \caption{Apple TV Non-trackers}
        \label{fig:3x3_apple_3}
    \end{subfigure}

    \stepcounter{row}
    \begin{subfigure}{0.32\textwidth}
        \includegraphics[width=\linewidth]{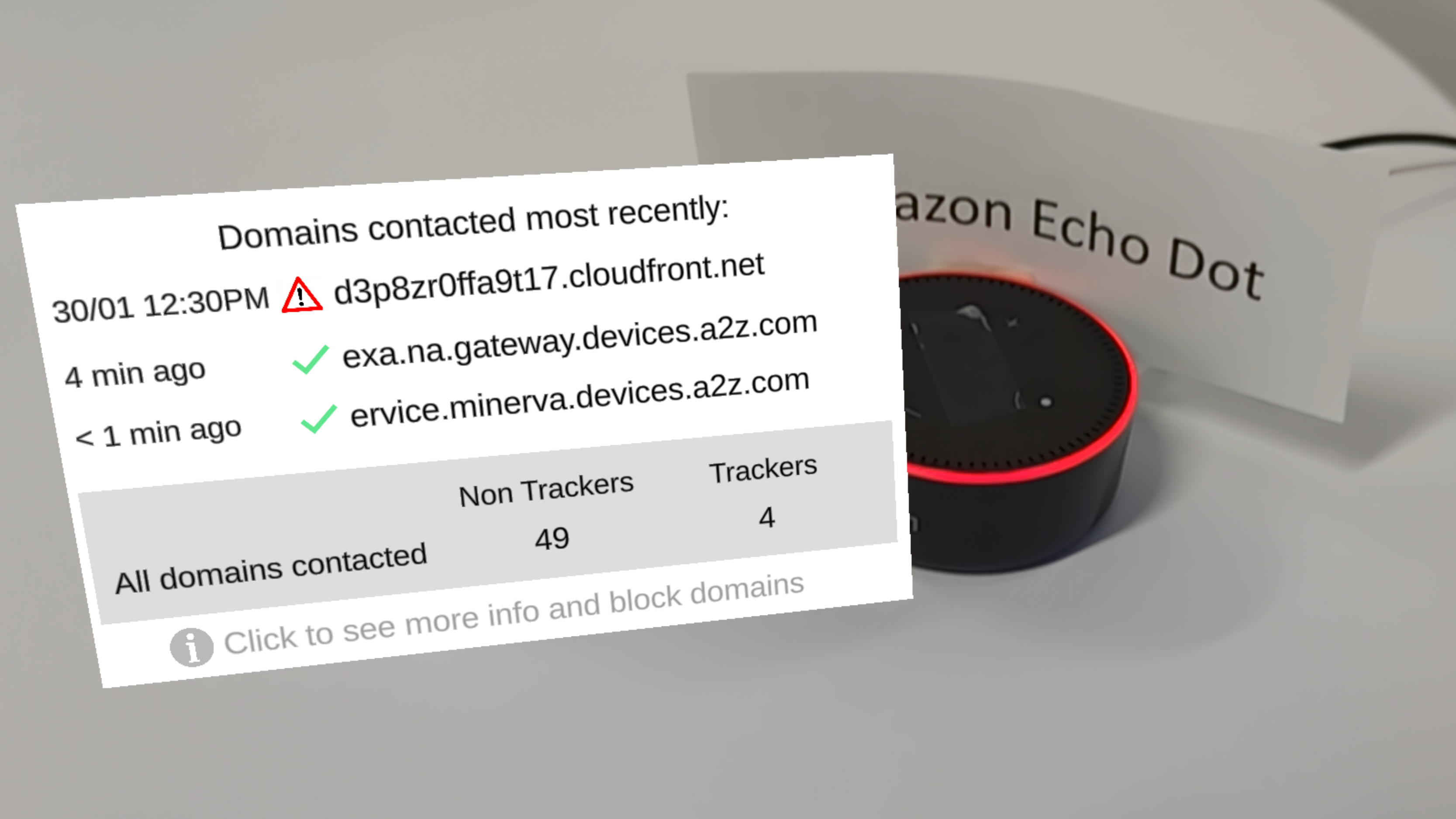}
        \caption{Echo Dot Recent Information}
        \label{fig:3x3_echo_1}
    \end{subfigure}
    \hfill
    \begin{subfigure}{0.32\textwidth}
        \includegraphics[width=\linewidth]{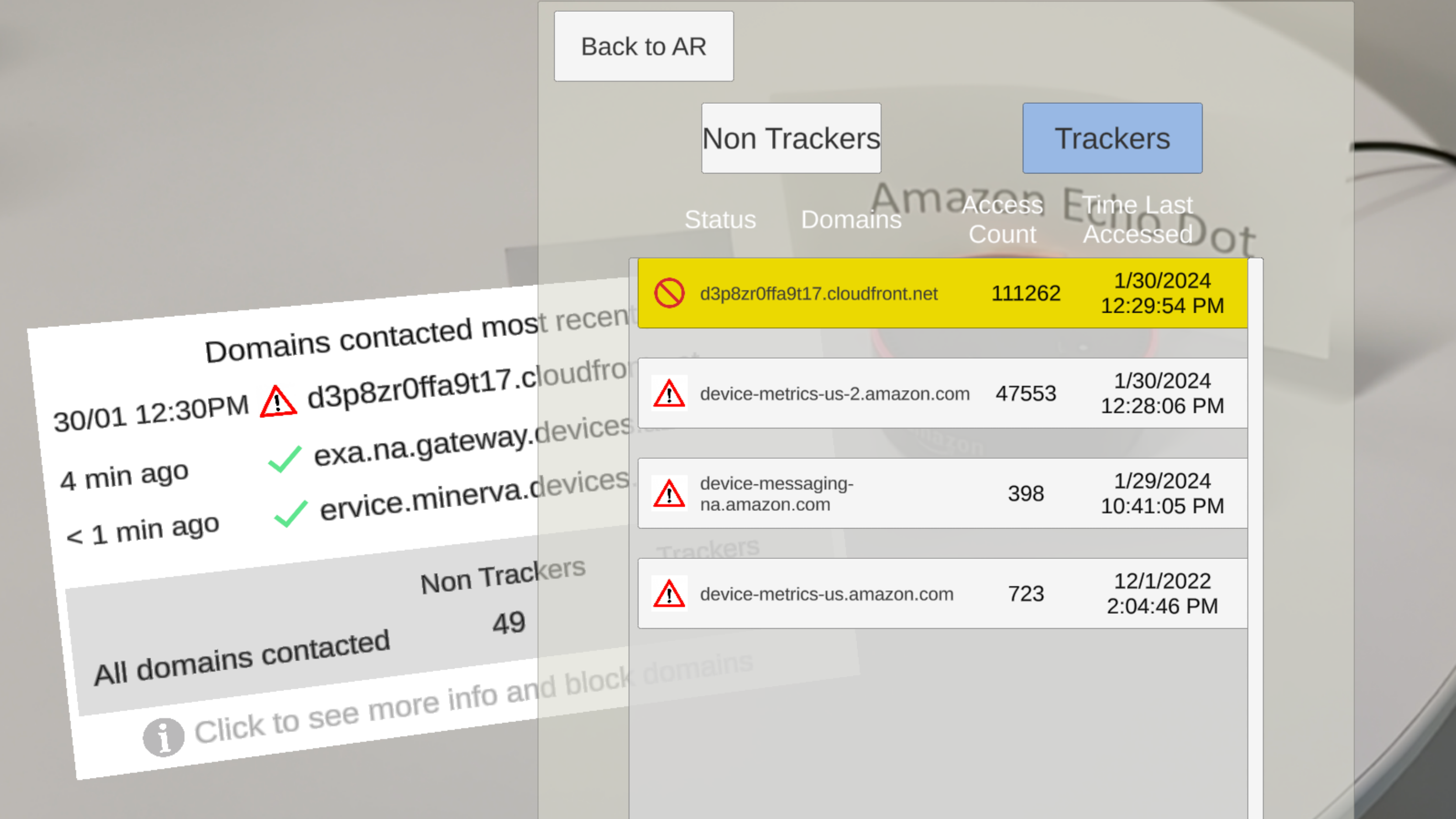}
        \caption{Echo Dot Trackers}
        \label{fig:3x3_echo_2}
    \end{subfigure}
    \hfill
    \begin{subfigure}{0.32\textwidth}
        \includegraphics[width=\linewidth]{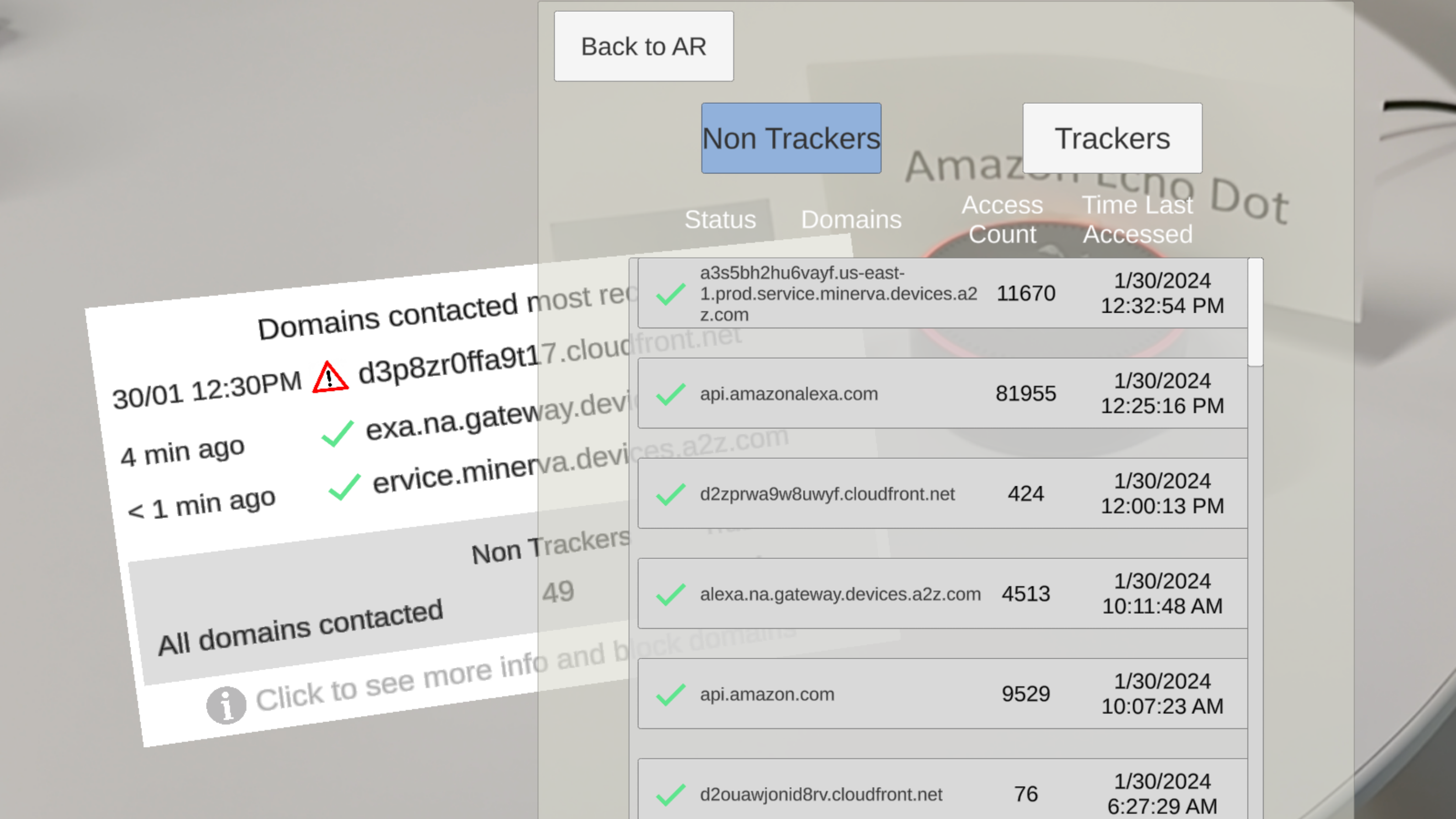}
        \caption{Echo Dot Non-trackers}
        \label{fig:3x3_echo_3}
    \end{subfigure}

    \stepcounter{row}
    \begin{subfigure}{0.32\textwidth}
        \includegraphics[width=\linewidth]{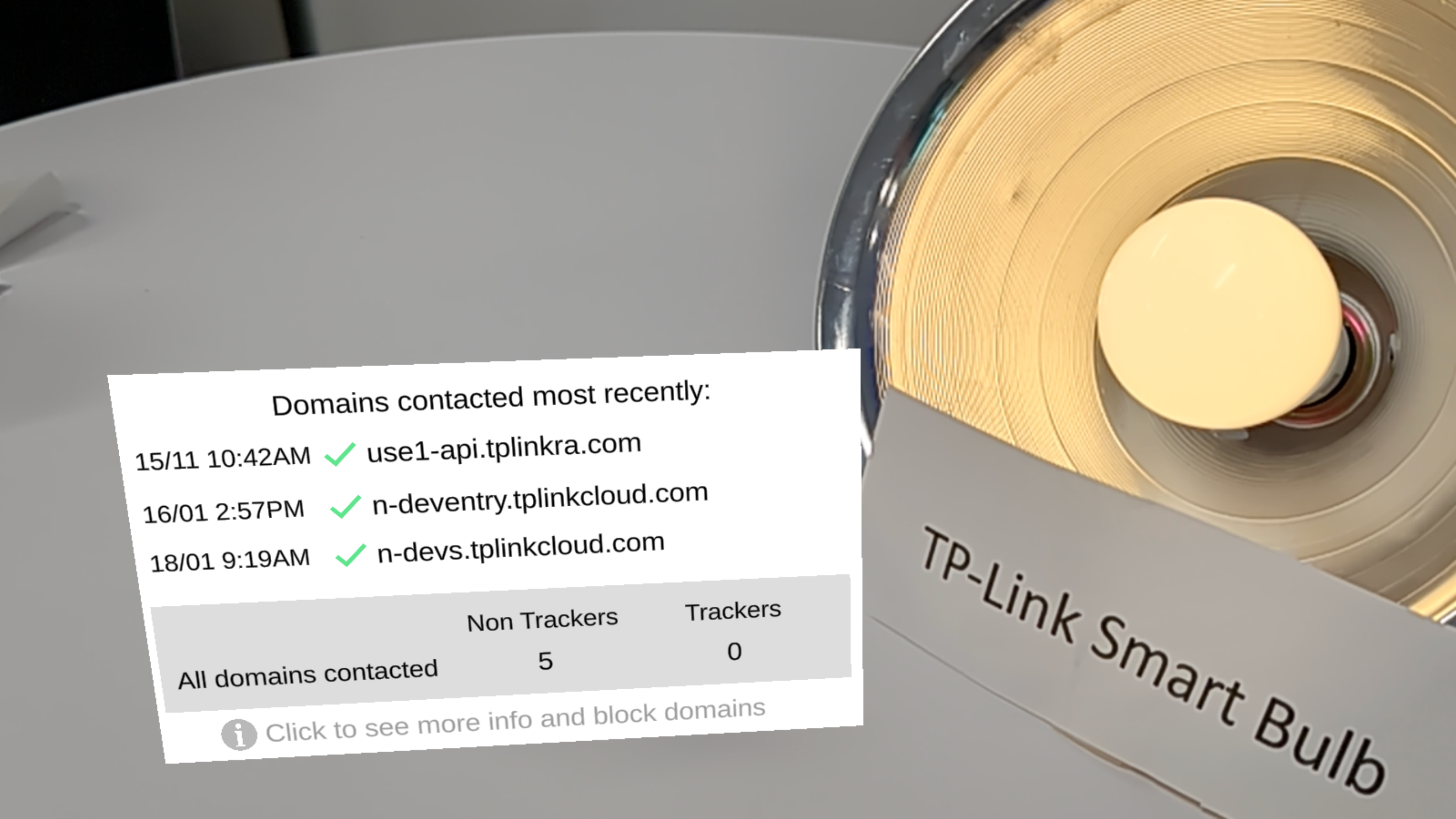}
        \caption{TP-Link Bulb Recent Information}
        \label{fig:3x3_bulb_1}
    \end{subfigure}
    \hfill
    \begin{subfigure}{0.32\textwidth}
        \includegraphics[width=\linewidth]{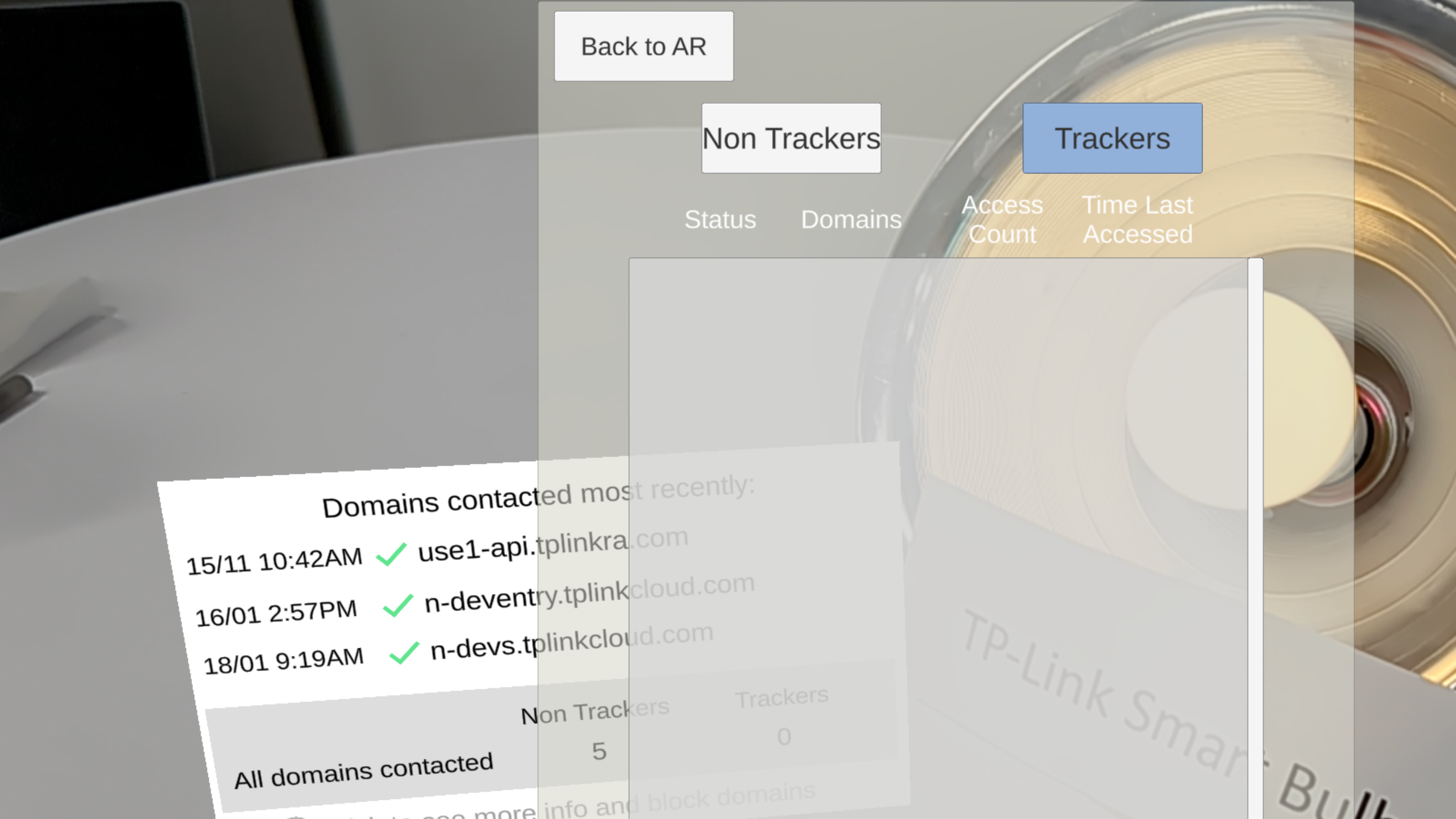}
        \caption{TP-Link Bulb Trackers}
        \label{fig:3x3_bulb_2}
    \end{subfigure}
    \hfill
    \begin{subfigure}{0.32\textwidth}
        \includegraphics[width=\linewidth]{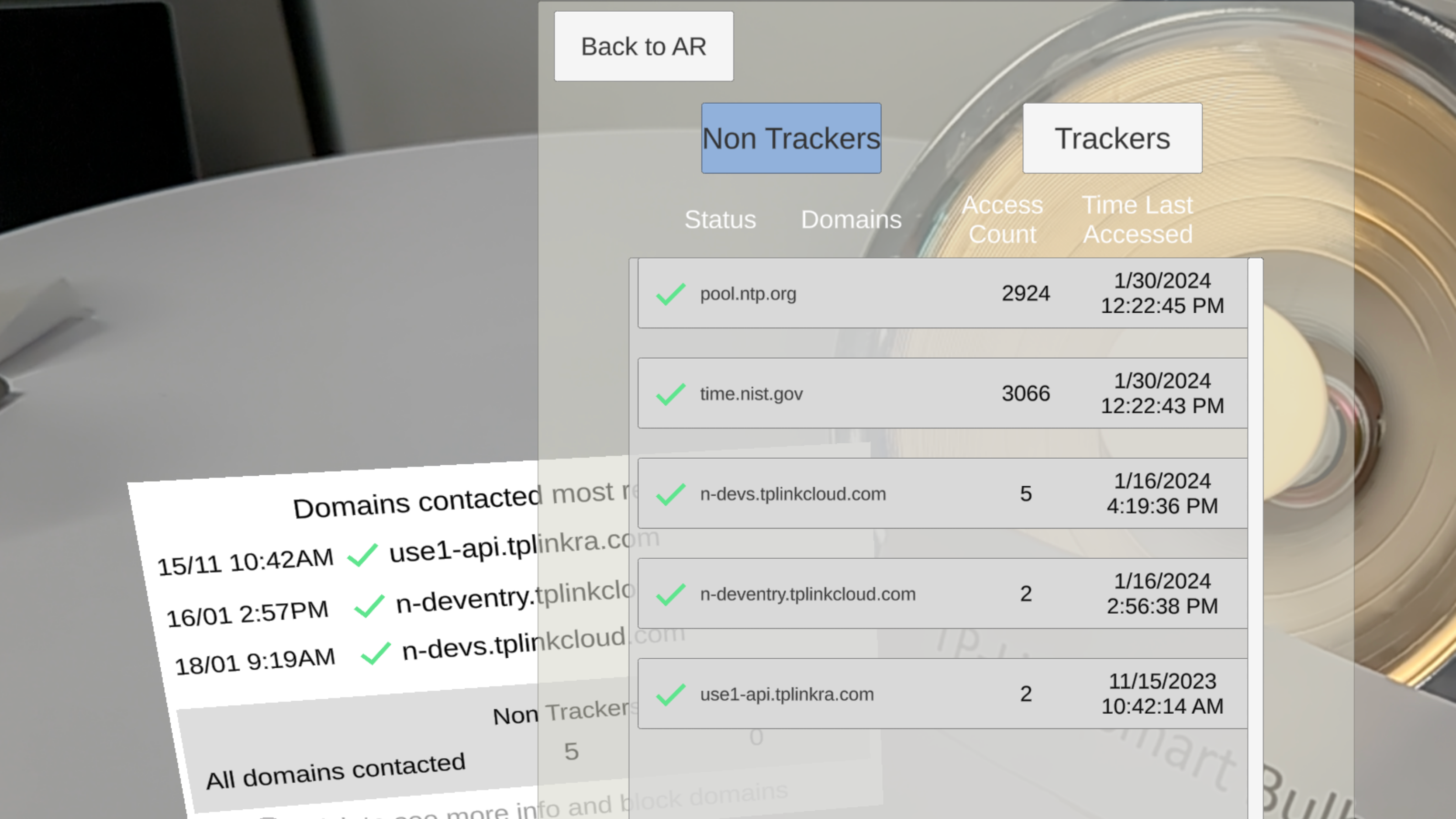}
        \caption{TP-Link Bulb Non-trackers}
        \label{fig:3x3_bulb_3}
    \end{subfigure}
    \captionsetup{font={footnotesize}}
    \caption{\tool{}'s Mixed Reality app. We show screenshots of \tool{}'s MR app for \textbf{(a)} Apple TV, \textbf{(b)} Amazon Echo Dot, and \textbf{(c)} TP-Link smart bulb. For each of the devices we show the following:
    \textbf{(i)} Recent Information Panel, overlaid in the environment once the marker is detected;
    \textbf{(ii)} Global Information Panel (appears when Recent Information Panel is clicked on) displaying tracker domains list 
    with the option of blocking domains, with blocked domains highlighted with yellow color and a red block symbol; and
    \textbf{(iii)} Global Information Panel displaying non-tracker domains list.
    }
    \label{fig:UI_flow_chart}
\end{figure*}
\fi

\subsection{Mixed Reality Mobile App}\label{subsec:mobile_app} 
The MR app displays real-time IoT device traffic information to the user by overlaying the information on top of the IoT device in Augmented Reality (AR) (see \autoref{fig:UI_flow_chart}). The app was built in Unity Engine~\cite{Unity} using AR Foundation~\cite{AR_Foundation}, AR Core~\cite{AR_Core}, and OpenCV~\cite{OpenCV}. It was tested on a Pixel 6 Android smartphone.

\subsubsection{\textbf{Mixed Reality Mobile App Design}} 
When the phone's camera is pointed at an IoT device's marker, the ``Recent Information Panel,'' shown in Figures~\hyperref[fig:3x3_apple_1]{2(i)}, pops up and is the first part of the MR experience that users interact with. 
The virtual panel floats in augmented reality near the device, such that the panel feels part of the real world.
We designed this panel to provide a brief overview of the latest IoT traffic, 
including how long ago the three most recent domains were contacted and sums of all of the trackers and non-trackers that particular IoT device has contacted. Non-tracker domains are represented with green check marks and trackers are represented with a warning symbol that looks like a red triangle containing an exclamation mark (\autoref{fig:3x3_echo_1}). 

In addition to this brief display, we sought to provide users with the option to view all of the network traffic data and to provide users the option to block unwanted traffic. The ``Recent Information Panel'' acts as a button itself, with text at the bottom of the panel instructing the user to ``Click to see more info and block domains.'' The click is made possible by a ``raycast'' from the mobile app screen space into the virtual 3D world space. The user’s touch on the screen is captured, and a ray is cast through the touch straight into the scene. In the event that the Recent Information Panel is clicked (i.e., the ray collides with the object and produces a collision), the ``Global Information Panel'' appears in the screen space. 
The ``Global Information Panel'' informs users about how frequently and how recently trackers and non-tracker domains are contacted to support users in making informed decisions about blocking trackers. 
The “Global Information Panel” contains lists of all of the non-trackers and trackers that the IoT device has contacted. For both non-trackers and trackers, the information shown includes domain name, domain contact frequency, timestamp of the last time the domain was contacted, and status symbol marking the domain as blocked (red circle with a slash) or not blocked (red triangle warning symbol). 
Users can block any of the tracker domains by clicking on a domain (see Figures \ref{fig:3x3_apple_2} and \ref{fig:3x3_echo_2}). When a domain is blocked, the row appears in yellow and the status symbol changes into a red circle with a slash through it. 

Interactions with the AR visualizations itself, both clicking on virtual objects and those interactions taking effect in the real world (e.g., blocking a domain), turn an AR experience into an MR experience, since reality is not only augmented with visualization but also interactive, as defined by Speicher \etal{}~\cite{speicher_what_2019}.

\subsubsection{\textbf{IoT Device Detection}}
The current \tool{} prototype uses ArUco markers \cite{garrido2014automatic} to  differentiate between IoT devices and to signal to the app where to place the visualization (see \autoref{appendix:UI_flow_chart_step1} in Appendix~\ref{app:mobile_app}). We chose to use markers since markers produce a quick and accurate 3D pose when detected. Markers continue to be widely used in AR and MR applications across various domains in Human-Computer Interaction and are popular due to their ease of integration and affordability~\cite{infraredtags_dogan_2022, stackable_music_chen_2023, ARwithmarker_app_2022, qruco_jenss_2023, dynatags_scheirer_2022}.
Once a marker is detected, and the pose of the marker is computed, the “Recent Information Panel” visualization is overlaid virtually,
placed in the virtual world with respect to the camera.
The panel is then anchored in the 3D virtual world space. That is, if the camera turns away from the marker and looks back at the IoT device again, the panel will remain floating in 3D, even if the marker is no longer detected. 

Marker-less object detection is out of scope for this work and orthogonal to the main contribution. Future versions of VBIT will consider different device detection methods, including 
marker-less detection, the devices announcing themselves, discovery, and in-door localization based on their traffic and signal~\cite{sharma_lumos_2022}.

\subsection{Interactive Web App}\label{subsec:web_app}

The web app provides more granularity, with a variety of graphs and more information. The app updates its contents in real-time by connecting to sockets on the server.
It was developed using Flask~\cite{Flask} and D3v4~\cite{d3v4} for generating  graphs. 

\subsubsection{\textbf{Web App Design}}
The web app has two main pages.
The Dashboard page presents an overview of metrics and data visualizations that enable users to monitor and analyze real-time information at a glance, and the device page provides detailed information on individual IoT devices connected to the system.

\ifanswers
\begin{figure}[t!]
\centering
\frame{\includegraphics[width=0.95\linewidth]{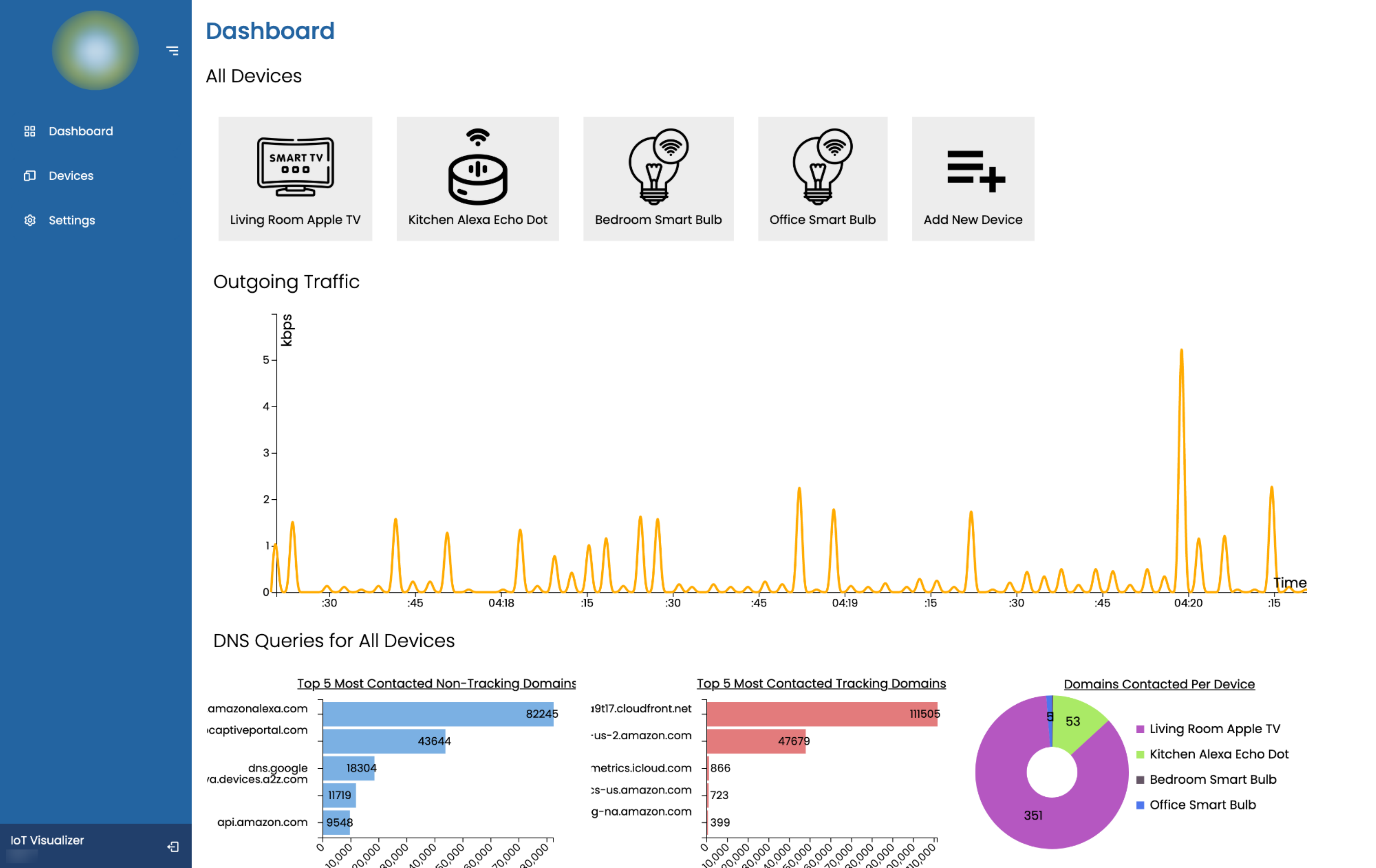}}
\caption{Dashboard of \tool{}'s Interactive Web App.}
\label{fig:dashboard_page}
\end{figure}
\fi

\parheading{\textbf{Dashboard Page.}}
The dashboard tab (\autoref{fig:dashboard_page}), displays aggregate data collected from all the IoT devices at the top of the page. First, the Outgoing Traffic line graph shows the Kilobits per second of the outgoing traffic from all IoT devices, which highlights if the IoT devices are sending data, how frequently they are sending it, and the size of the data being sent. The spikes in the outgoing traffic line graph may reveal potential IoT device background communication occurring when IoT devices are not in use. The user can then check the specific device that caused this spike and block any tracker domains that were contacted.
Second, two bar graphs show the top five tracker and non-tracker domains contacted for all devices, and a pie chart displays domains contacted per device. For example, the most contacted non-tracker domain is "amazonalexa.com" in \autoref{fig:dashboard_page}. These figures help users identify tracker domains to block and devices that communicate the most.
Third, under the bar graphs and pie chart, an alluvial diagram (not depicted in the \autoref{fig:dashboard_page}) 
maps the IoT devices $\rightarrow$ Second-Level Domain (SLD) $\rightarrow$ Organization $\rightarrow$ Trackers or Non-trackers. This diagram helps the user understand the breakdown of domains per device and the organizations behind these domains. Users may choose to refrain from sharing their information with a specific organization by blocking domains linked to that organization.

\parheading{\textbf{Devices Page.}}
This page has the same information shown in the MR app, in addition to a pie chart comparing the number of unblocked/blocked trackers to non-trackers, an alluvial diagram for a selected device (\autoref{app_fig:devices_alluvial} in Appendix~\ref{app:web_app}) similar to the one for all devices on the Dashboard page, and a line graph showing the number of DNS queries sent over time.

\subsection{Performance Evaluation}\label{subsec:performance}
\subsubsection{\textbf{Network Performance Evaluation}}

In order to assess the efficiency of the network traffic interception and blocking module,
we ran similar evaluations and compared our results with Privacy Plumber~\cite{ndss_2023_ar_privacy} since it is the closest system to \tool{}. Unlike Privacy Plumber, \tool{} does not have any major source of bottleneck since it does not rely on ARP spoofing. \tool{}'s components communicate on a per-need basis.
We installed the Ookla Speed Test app~\cite{ookla} on an Android phone connected to \tool{}'s Wi-Fi network. We ran 15 back-to-back speed tests to measure the inbound and outbound traffic rates for three scenarios. Without \tool{}, the average inbound and outbound rates were 15.48 and 21.88 Mbps, respectively. With \tool{} running, the average rates were 15.37 (0.37\% decrease) and 21.87 (0.046\% decrease) Mbps, respectively. With \tool{} running and handling requests to block and unblock traffic from both MR and web apps, the rates were 15.14 (2.2\% decrease) and 21.54 (1.55\% decrease) Mbps, respectively. Thus, the network traffic overhead of running \tool{} is negligible. In comparison, Privacy Plumber reduced the inbound and outbound rates by 85.9\% and 22.6\%, respectively. See Appendix~\ref{app:power_rpi} for power consumption measurements.

\subsubsection{\textbf{App Performance Evaluation}}
We measured the battery lifetime for the MR app using the AccuBattery app~\cite{accubattery} on the Android phone that ran \tool{} and measured its power consumption \wrt{} battery lifetime. Additionally,
we measured the power consumption of the Pixel Camera app for comparison purposes because it also uses the camera the entire time it is running.
With only the app of interest running, we ran the measurement for 10 minutes for each app. \tool{}'s MR and the Pixel Camera apps consumed 37.5\%/h (193.9 mAh) and 22.1\%/h (109.2 mAh), respectively.
The power consumption of the MR app is higher than the Pixel Camera app
likely because \tool{}'s MR app, in addition to using the phone's camera, frequently communicates with Pi-hole to update the list of trackers displayed for each device and performs other computer vision background processing, \eg{} tracking the phone's pose in 3D space and searching for markers in each frame. 
This should not be an issue since we foresee that users would typically use the MR app
briefly for checking and blocking/unblocking traffic.

%% file: sections/04-study_methods.tex
\section{User Study Design}\label{sec:user-study}
To evaluate \tool{}, we conducted a survey with 200 participants and in-person semi-structured interviews with 18 participants. Both study protocols were reviewed by our institutional review board (IRB) and were deemed exempt. Regarding ethical considerations, our user study did not collect personally identifiable information, and we informed participants that their participation was voluntary and that they could withdraw at any time without penalty. The survey and interview were designed to capture participants' perceptions about IoT privacy before and after learning about \tool{} and to evaluate the usability of \tool{}. We conduct both survey and interview because the survey enables us to analyze at scale the usability of \tool{} and factors that influence users' perceptions, which is complemented by the interviews that allow participants to use \tool{} in person and provide in-depth discussion. 

Our study and analyses are guided by four user study questions (UQ) listed below, where UQ0 is aimed at better understanding our participants' perceptions and expectations about IoT privacy before learning about \tool{} and UQ1-3 investigate participants' perceptions and expectations after learning about \tool{}. UQ0 has been explored in prior work, as discussed in Section~\ref{sec:relatedwork}, but we still include UQ0 in our study design to compare our participants' attitudes before and after learning about \tool{}. We recruited participants with varying technical knowledge and background in privacy to evaluate \tool{} with a wide range of users, and thus UQ0 also provides context about our participants.

\begin{itemize}
    \item[\textbf{UQ0:}] Before learning about \tool{}, are users \emph{aware} of and/or \emph{concerned} about privacy of IoT devices \wrt{} advertising and tracking? What \emph{factors} influence users' perceptions, preferences, and actions regarding IoT advertising and tracking?
    
    \item[\textbf{UQ1:}] How does \emph{\tool{} influence} users' perceptions, preferences, and actions regarding IoT advertising and tracking?
    
    \item[\textbf{UQ2:}] How \emph{usable} is \tool{} currently and what factors influence users' perceptions of \tool{}?
    
    \item[\textbf{UQ3:}] How can \tool{} be \emph{improved} to give better user experience? 
    
\end{itemize}

\subsection{Survey}

\subsubsection{\textbf{Participants}}
We built our survey using Qualtrics~\cite{qualtrics} and recruited 200 participants on Prolific~\cite{prolific}. Our participants were required to be 18 years or older. We did not limit the recruitment on Prolific to any specific technical background, as our goal is to evaluate \tool{} with users from a wide range of backgrounds. To ensure data quality, we only allowed participants with at least a 95\% approval rate on Prolific to participate, and each participant was only allowed to submit the survey once. We checked for invalid responses such as incomplete responses or those that failed our attention check and did not need to remove any responses. The survey questionnaire is included in Appendix~\ref{appendix:questionaire}.

Among the 200 participants, 47.5\% are male, 49.0\% are female, 2\% are non-binary, and 1.5\% preferred not to answer. Our participants were mostly young adults (37.0\% were between 25 and 34 years old) and highly educated (88.0\% completed at least a college degree). See Table~\ref{app_tab:demographic_survey} in Appendix~\ref{app:demographics} for the complete demographic information.

\subsubsection{\textbf{Survey Protocol}}
The survey included 35 questions and took approximately 20 minutes to complete. We compensated participants at a payment rate of \$12/hour recommended by Prolific. 
Participants were prompted to review the study information sheet and give consent to participate. The survey included multiple-choice, open-ended, and Likert-scale questions. We collected demographic information (Q2-Q4) and then began the pre-demo background questions (Q5-Q12), including questions about participants' experiences and opinions regarding IoT devices, targeted advertising, and ad blockers. 
Next, the \tool{} demo portion (Q13-30) presented participants with videos of the MR and web apps, each followed by questions to evaluate their usability, such as which components were most important to users and suggestions for improvements. 
The post-demo portion (Q31-33) included questions about whether users had learned any new information as well as questions similar to the pre-demo portion to measure any changes after learning about \tool{}, such as their willingness to install \tool{} and level of concern \wrt{} IoT privacy.

\subsubsection{\textbf{Quantitative Data Analysis}}
We conducted quantitative data analysis on the survey results to analyze differences and trends between groups as well as opinions before and after the demo videos. We used the Kruskal-Wallis test to analyze our non-parametric data, including responses to multiple-choice and Likert-scale questions. Statistical significance is assumed for $p\leq0.1$, which we indicate with bold text in our result tables in the later sections. We denote highly significant results ($p\leq0.05$) with `*'. We also conducted the post-hoc Dunn's test for pairwise comparisons of independent variables.
Responses to open-ended questions were independently coded by two researchers in our team using a codebook. New codes were added to the codebook throughout the process and responses were re-coded after codebook updates. We ensured inter-rater agreement by discussing and finalizing the codes together.

\subsection{Interviews}

\subsubsection{\textbf{Participants}}
We recruited interview participants via email and flyers. Interested participants were invited for an in-person interview session with our team. Similarly to the survey study, we did not limit the recruitment to participants of any specific technical background.
Among the 18 participants, 55.6\% are male, 38.9\% are female, and 5.6\% preferred not to answer. All participants were between 18 to 34 years old. See Table~\ref{app_tab:demographic_interview} in Appendix~\ref{app:demographics} for the complete demographic information.

\subsubsection{\textbf{Interview Protocol}}
We conducted 18 in-person interviews in our lab with \tool{} set up for hands-on exploration. All interviewees provided consent to participate and to have the audio of the interview recorded. Interview sessions lasted approximately 43 minutes on average, ranging from 30-60 minutes for all interviewees. We compensated each interviewee with a \$20 gift card. 
The interviews were similar to the survey \wrt{} structure and questions, but we conducted the interviews in a semi-structured manner, allowing for in-depth discussion and app exploration. In the \tool{} demo portion, we explained briefly how each app worked and invited the interviewees to explore and talk us through what they were understanding (or not understanding) about \tool{}. This process allowed us to analyze the level of difficulty in navigating and understanding each app for first-time users, and then we explained how the apps worked and clarified any confusion.
Additionally, we collected demographic information and conducted the SUS questionnaire (Appendix~\ref{appendix:interview_sus}) for the MR and web apps separately.

\subsubsection{\textbf{Qualitative Data Analysis}}
We generated transcripts from the interview recordings using Microsoft Word's AI-based transcription feature (R3), and then two researchers in our team manually verified and corrected the transcripts. We conducted iterative thematic analysis~\cite{Braun2019}, in which two researchers open-coded each participant's transcript independently before merging and finalizing the codes together through discussions. The researchers discussed the results to come to an agreement on the key emerging themes.

%% file: sections/05-results.tex
\section{User Study Results}\label{sec:results}
This section reports the findings from the survey and interviews. 
In Section~\ref{sec:participant_background}, we provide context about the participants' privacy perceptions and factors that influence them before learning about \tool{} (UQ0).
Then, we report the key themes that emerged from our user study regarding the usability and design of \tool{} (UQ1-3).

Section~\ref{sec:results_vbit_impact} discusses \tool{}'s impact on participants' privacy perceptions (UQ1), and 
Section~\ref{sec:results_vbit_usability} reports on \tool{}'s usability scores (UQ2). 
Finally, Section~\ref{sec:results_vbit_design} discusses participants' suggestions for improving \tool{}, highlighting their priorities regarding transparency, control, and customizability (UQ3).
We indicate quotes from the survey and interviews with `S' and `I', respectively.

\subsection{Participants' Privacy Background \& Perceptions (UQ0) \label{sec:participant_background}}

\subsubsection{\textbf{Participants' Privacy Concerns}}
Participants had varying levels of concern regarding ads and tracking in IoT devices, which aligns with previous research~\cite{user_iot_concerns_zeng_2017, smart_home_privacy_perceptions_zheng_2018, spaf_das_2023}. While many experienced targeted ads and believed their activities were tracked, some felt uncomfortable and avoided these devices, whereas others were willing to accept the tracking due to its perceived utility or inevitability. We provide further details regarding participants' privacy concerns regarding ads and tracking in general and specifically with respect to IoT devices in Appendix~\ref{app:survey_results}. We asked this set of questions simply to gauge the participants and check whether their opinion matches previous results or not.

\subsubsection{\textbf{Participants' Preferences}}
All interviewees wanted to learn more about trackers and desired transparency and control over their IoT devices. We asked interviewees what details about trackers they wanted and what features they would want in an IoT ad/tracker blocker in the pre-demo portion. Most interviewees (15/18) wanted to learn about how trackers work, such as the collection, sharing, and processing of user data, how trackers can track users across services and apps, and how to block trackers. Two interviewees also wanted to figure out how to avoid privacy-invasive websites with large numbers of trackers or websites that engage in non-private data sharing. I2 shared that they wanted to learn more about trackers to be able to block them: \textit{``First of all, how they got my information, and if possible, if there are ways I could circumvent that to avoid handing out unnecessary information to other people.''} Three interviewees were not sure what details they would want to learn, as they were unfamiliar with trackers.

Regarding desired features for an IoT ad/tracker blocker, interviewees mentioned ease of installation, usability, control over trackers, flexibility in blocking, and transparency over their IoT devices' behaviors. Control and transparency were the most common responses, with 17/18 interviewees mentioning these kinds of features, demonstrating the need for a system that provides these functionalities to empower users and enable more informed decision-making, as I14 explains: \textit{``Something that I can choose. I would still like to give some data out because that will help me in return, but I would like to know what I’m sending, who I’m sending to. I should have control, like disable at any point and change it. I should know at least what I’ve accepted so far also.''}

\subsubsection{\textbf{Factors Influencing Users' Perceptions}}
\ifanswers
\begin{figure}[t]
\centering
\includegraphics[width=0.85\linewidth]{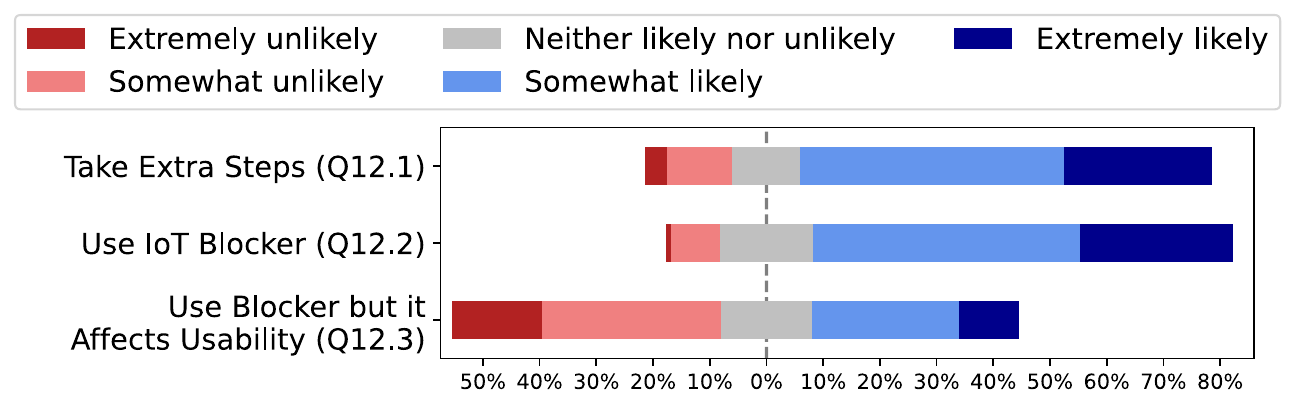}
\caption{Participants' likelihood to use an ad blocker.}
\label{fig:use_blocker}
\end{figure}
\fi 

Usability and the types of IoT devices that participants own are significant factors that influence their perception of IoT privacy and decisions to detect/block IoT trackers.

\parheading{Usability:}
Participants are likely to detect/block trackers on IoT devices, but the usability trade-off matters. Over half ($61.5\%$) of survey participants have used an ad blocker, where the majority of those are plugins installed on a browser ($89.43\%$). As shown in Figure~\ref{fig:use_blocker}, over $70\%$ of participants would take extra steps to detect/block trackers on IoT devices. However, only $36.5\%$ of users would use an ad blocker if it affected the usability of the device.

Most interviewees (14/18) reported using browser/mobile ad blockers primarily to block ads and improve their user experience online or to stop tracking.
Interviewees who did not use ad blockers (4/18) felt they were unnecessary, either because they have grown accustomed to ads or do not see enough ads to warrant it. Regarding ads on YouTube that users cannot skip, I1 shared, \textit{``At this point, I’m just used to it, to be honest."}
Additionally, a majority of interviewees (15/18) were interested in installing an IoT ad/tracker blocker for increased control and transparency of their IoT devices, but some (3/18) seemed unwilling because of concerns over the trustworthiness of such a system or because they felt it was unnecessary to block ads/trackers on IoT devices, as I6 explained: \textit{``I think there is no need to use those.”}

Many interviewees (10/18) were willing to tolerate some breakage or delay caused by an ad blocker to improve their user experience but would deactivate the ad blocker if it caused too much disruption. A few interviewees (3/18) were willing to sacrifice usability more than others to protect their privacy, as I3 explained: \textit{``It may affect the function or performance of the device, but I prefer with the ad blocker."} On the other hand, the rest (5/18) would only tolerate imperceptible delays, prioritizing their usability over their privacy. I2 shared, \textit{``As annoying as the ads are, I think my workflow not being interrupted is my first and foremost priority."}

\parheading{IoT Devices Owned:}
The type of IoT devices users own affects their privacy comfort and preferences regarding ads/tracking.
Table \ref{tab:p-values-tables-iot-correlation} contains the significant values for the top five IoT devices owned by our survey participants (as shown in Table \ref{app_tab:iot_devices_owned} in Appendix~\ref{app:survey_results}).

Users of Smart TVs believe that IoT devices track their daily activities more than others. This could be because Smart TVs have more ways to display ads, and thus users are exposed to them at a higher rate, leading to concerns of tracking. The domain information shown also tends to be more important, which might be explained by the importance of knowing who is being contacted by these Smart TVs.
For similar reasons as the Smart TV users, voice assistant users tend to rate the list of trackers with higher importance, would more likely install \tool{}, and are more concerned about privacy after seeing the \tool{} system. These results might be due to the microphone in voice assistants that is constantly listening to wake words, which could lead users to think that the device is recording everything being said.
Additionally, users have more privacy comfort when using smartwatches and smart bulbs, which may be because these devices seem less intrusive or more controllable than other IoTs.

\begin{table}[t]
    \footnotesize
    \centering
    \caption{The type of IoT device owned affects user perceptions and preferences (Kruskal-Wallis test).}
\begin{tabular}{ p{0.25\linewidth}p{0.4\linewidth}p{0.15\linewidth} }
\specialrule{.1em}{0em}{0em}
\textbf{Independent\newline Variable} & \textbf{Dependent Variable} & \textbf{p-value} \\
\specialrule{.1em}{0em}{0em}
\multirow{2}{=}{Smart TV}& IoT Distrust & \textbf{0.055} \\
& Domain Information Shown & \textbf{0.034*} \\
\specialrule{0.05em}{0em}{0em}
\multirow{3}{=}{Voice Assistant}& List of Trackers (MR) & \textbf{0.007*} \\
& Would You Install \tool{} & \textbf{0.035*} \\
& Privacy Comfort (Post Videos) & \textbf{0.082} \\
\specialrule{0.05em}{0em}{0em}
Smartwatch & Level of Privacy Comfort & \textbf{0.049*}\\
\specialrule{0.05em}{0em}{0em}
Smart Bulb & Level of Privacy Comfort & \textbf{0.028*} \\
\specialrule{0.05em}{0em}{0em}
Smart Thermostat & \multicolumn{2}{c}{No significance} \\
\specialrule{.1em}{0em}{0em}
\end{tabular}
\label{tab:p-values-tables-iot-correlation}
\end{table}

\subsection{\tool{}'s Impact on Users (UQ1)}\label{sec:results_vbit_impact}
\subsubsection{\textbf{\tool{} Effect on Awareness}}
VBIT helped educate participants and raise their awareness about tracking in IoT. The majority of the survey participants (92.5\%) and all interviewees reported that they learned new information or gained awareness about IoT ads and tracking from \tool{}. Such new information included IoT device communication (\eg{} that IoT devices contact domains, what domains are), what trackers are, the ability to block domains, or that a system to detect/block trackers like \tool{} is possible. Most interviewees (15/18) were shocked by the number of trackers contacted by IoT devices and the frequency of tracking they observed while exploring \tool{}. Figure~\ref{app_fig:new_info_learned} in Appendix~\ref{app:survey_results} summarizes the information that the survey participants considered to be new to them. Many survey participants did not know the extent of tracking found in IoT devices, as S182 shared: \textit{``I sort of already knew just how bad the tracking was but actually seeing the trackers was a bit unsettling. It's like the old saying: if something is free, then you are the product. But at least with this service, we can protect our privacy a bit more.''} I16 explained that increasing awareness encourages users to install a tracker blocker as a result: \textit{``I need to get one of these for myself. It's great because I wouldn't have thought there would be this many that I needed to be aware of. So I think just putting that on people's radar and making you feel more motivated to have a tracker system established.''}

\subsubsection{\textbf{Willingness to Install \tool{}}}
We asked the participants if they would install an IoT ad/tracker blocker before and after learning about the \tool{} system, and we found high statistical significance ($p = 0.002$) based on the survey results, indicating that the survey participants were more willing to install an IoT ad/tracker blocker after learning about \tool{}. The results show the lower quartile decreased (interquartile range decreased, Figure \ref{fig:install_iot_blocker_opinion}) increasing the likelihood of installing an IoT ad/tracker blocker. However, the level of privacy comfort when using IoT devices did not change significantly after the VBIT demo.

\ifanswers
\begin{figure}[t]
\centering
\includegraphics[width=0.9\linewidth]{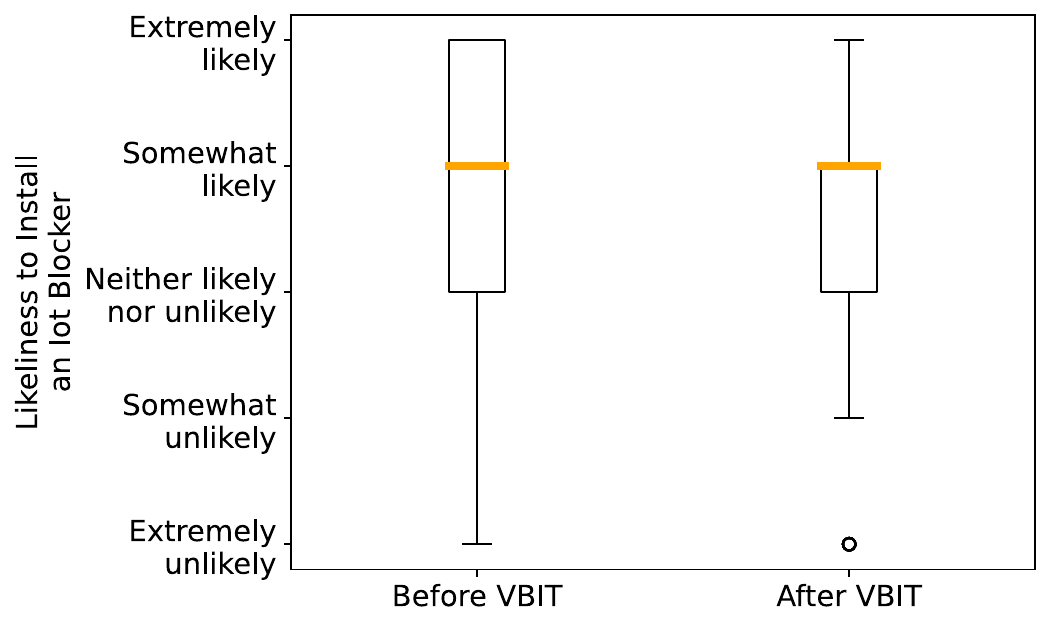}
\caption{Change of opinion regarding installing an IoT ad/tracker blocker after seeing the \tool{} system.}
\label{fig:install_iot_blocker_opinion}
\end{figure}
\fi

Similarly, all interviewees expressed that they would install \tool{} in their homes, demonstrating a change of opinion for participants who were originally unwilling to install such a system. Interviewees want to use \tool{} because it gives them control over their IoT devices through the blocking functionality and provides transparency through visualizations. I3, who avoids IoTs due to privacy concerns, shared that they may now even consider purchasing IoT devices if they can use  \tool{}: \textit{``I like it. I may possibly buy some smart devices because of this app. The reason why I don’t like smart devices is because they may record some other information without my permission. If this app exists, I may buy more.''} 
I18 expressed eagerness to install \tool{}, even if it required a substantial amount of effort: \textit{``It wouldn't affect, I think. I would, no matter what, still get it.''}
I6 shared that even though they may not choose to block trackers, they still want to install \tool{} as it gives them a sense of control and increased transparency: \textit{``Another thing is the transparency and also more in control because we bought these devices and it’s in our home, so I definitely want to know everything. Even though I might not block those trackers, it’s really good to know how many trackers it’s connected to and how many access count.''}

After the \tool{} demo, almost all interviewees (17/18) expressed that they do believe IoT devices pose privacy risks, such as security breaches and overall lack of transparency and control. However, their levels of concern \wrt{} using IoT devices remained the same as before the demos, similar to the survey results. Many interviewees (11/18) shared that they will still use their IoT devices because of their convenience. I7 reevaluated their IoT usage after the demo, expressing a desire for something like \tool{} to mitigate their concerns: \textit{``I should stop using them because they collect so much data. But, at this point, I think I’m too used to it. They are so much part of our lives. Unless there’s a better alternative which completely blocks off things, I feel like I’ll still use them.''}

\subsubsection{\textbf{Background Factors}}
We found several background factors that significantly affect survey participants' privacy perceptions and decisions toward IoT privacy.
We check for correlation between a list of factors (\eg{} demographics, IoT experience, ad blocker experience) and questions related to IoT privacy, willingness to install blockers, and new information learned. The p-values can be found in Table \ref{table:p-values-tables-correlation}.

\parheading{IoT Distrust:} There exists a correlation between IoT distrust and prior experience with IoT targeted ads. The more ads the participants experienced, the higher their distrust (Figure \ref{fig:iot_distrust}). This may be due to ads feeling intrusive, leading participants to believe that their IoT devices are tracking them.

\ifanswers
\begin{figure}[t]
\centering
\includegraphics[width=0.9\linewidth]{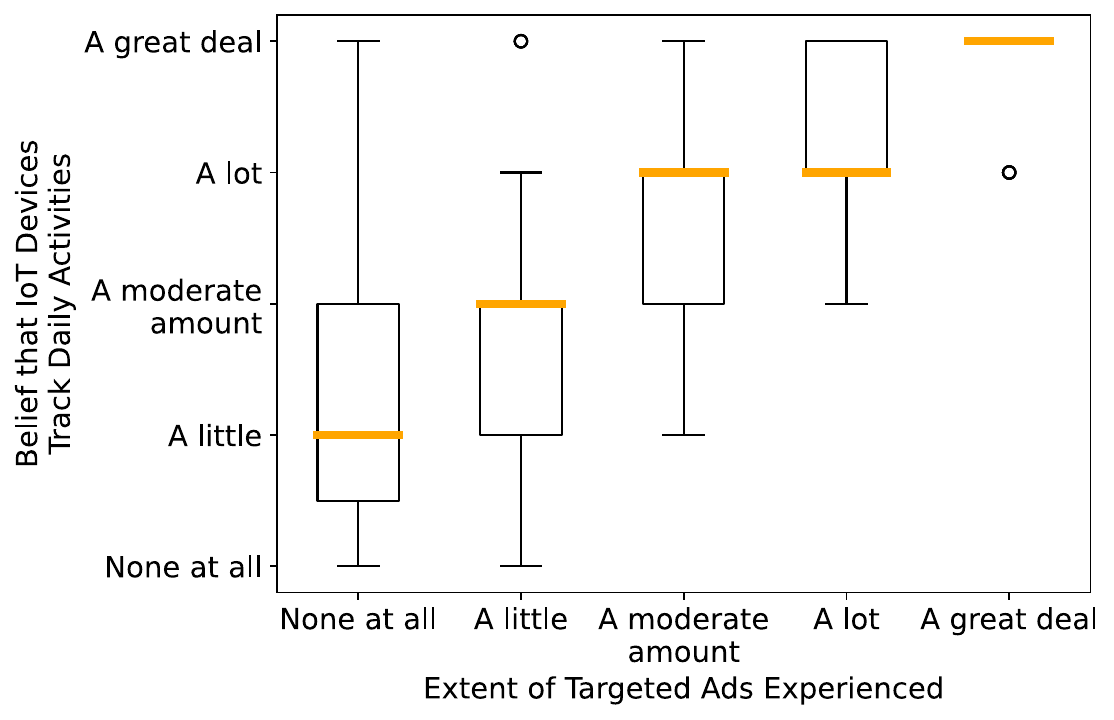}
\caption{Belief in IoT tracking by experience of targeted ads.}
\label{fig:iot_distrust}
\end{figure}
\fi

\ifanswers
\begin{figure}[t]
\centering
\includegraphics[width=0.75\linewidth]{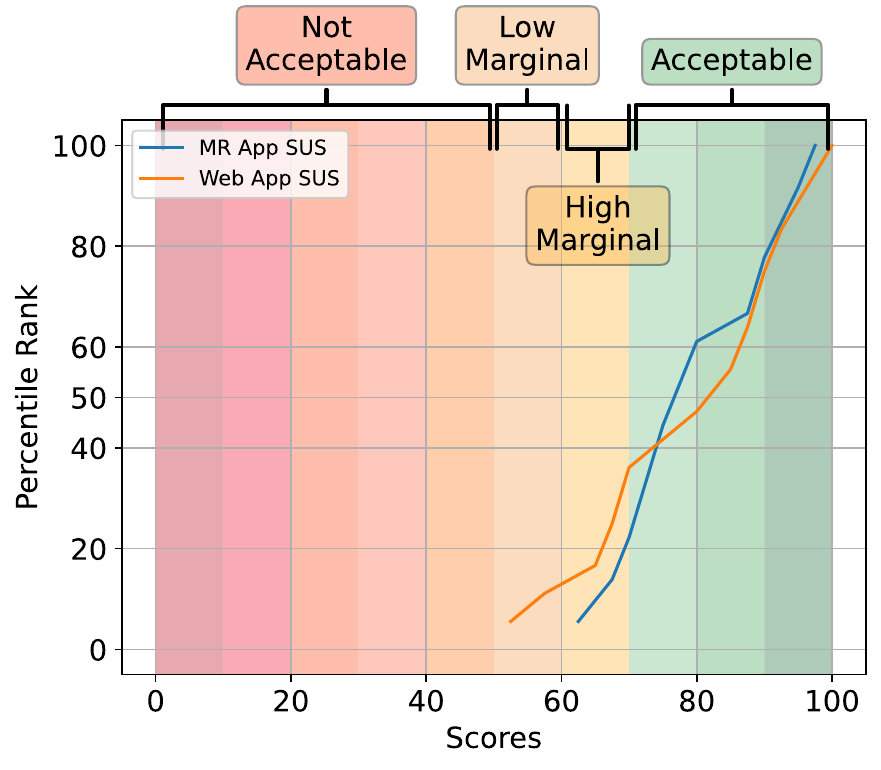}
\caption{Interview MR vs. web app SUS scores.}
\label{fig:mr_vs_web_SUS}
\end{figure}
\fi

\newcolumntype{R}[2]{
    >{\adjustbox{angle=#1,lap=\width-(#2)}\bgroup}
    l
    <{\egroup}
}
\newcommand*\rot{\multicolumn{1}{R{35}{1em}}}
\newcommand{\myTableWidth}{0.75}

\begin{table*}[t]
  \footnotesize
  \centering
  \caption{Difference of survey participant perception as a function of demographic and perception factors (Kruskal-Wallis).}
  \begin{tabular}{p{5cm}p{1.15cm}p{\myTableWidth cm}p{\myTableWidth cm}p{1.1cm}p{1.1cm}p{1.1cm}p{\myTableWidth cm}p{\myTableWidth cm}p{\myTableWidth cm}p{\myTableWidth cm}p{\myTableWidth cm}p{\myTableWidth cm}p{\myTableWidth cm}p{\myTableWidth cm}p{\myTableWidth cm}p{\myTableWidth cm}p{\myTableWidth cm}p{\myTableWidth cm}p{\myTableWidth cm}p{\myTableWidth cm}}
    \specialrule{.1em}{0em}{0em}
    \textbf{Independent Variable} & \rot{\textbf{IoT Distrust}} & \rot{\textbf{Privacy Comfort}} & \rot{\textbf{Used Blockers}} & \rot{\textbf{Extra Steps}} & \rot{\textbf{Would Use IoT Blockers}} & \rot{\textbf{Blocking vs Usability}} & \rot{\textbf{New Info Learned}} & \rot{\textbf{Would Install \tool{}}} \\
    \specialrule{.1em}{0em}{0em}
    Age \textbf{(Q2)} & 0.368 & \textbf{0.084} & 0.958 & 0.527 & 0.965  & 0.946 & 0.550 & 0.911\\
    \specialrule{0.05em}{0em}{0em}
    Education Level \textbf{(Q3)} & 0.513 & 0.802 & \textbf{0.021*} & 0.133 & 0.486 & 0.457 & \textbf{0.025*} & 0.612 \\
    \specialrule{0.05em}{0em}{0em}
    IoT Work Experience \textbf{(Q5)} & 0.299 & 0.244 & 0.219 & 0.164 & \textbf{0.075} & \textbf{0.016*} & 0.316 & 0.124 \\
    \specialrule{0.1em}{0em}{0em}
    Have you experienced targeted ads with IoT\newline devices? \textbf{(Q8.1)} & \textbf{<0.01*} & 0.269 & 0.113 & 0.106  & 0.393 & 0.945 & 0.257 & 0.253 \\
    \specialrule{0.05em}{0em}{0em}
    Do you think IoT devices track your daily activities to send you personalized ads/services?\newline \textbf{(IoT Distrust, Q8.2)} & - & \textbf{0.005*} & 0.425 & \textbf{0.002*} & \textbf{0.005*} & \textbf{0.065} & \textbf{0.058} & \textbf{0.047*} \\
    \specialrule{0.05em}{0em}{0em}
    Rate your level of privacy comfort when using IoT devices \textbf{(Privacy Comfort, Q9)} & - & - & 0.138 & \textbf{0.013*} & \textbf{0.006*} & \textbf{0.068} & 0.409 & 0.624 \\
    \specialrule{0.05em}{0em}{0em}
    Have you ever used any ads and trackers blockers? \textbf{(Used Blockers, Q11)} & - & - & - & \textbf{<0.01*} & \textbf{<0.01*} & \textbf{<0.01*} & 0.608 & 0.211\\
    \specialrule{.1em}{0em}{0em}
  \end{tabular}
  \label{table:p-values-tables-correlation}
\end{table*}

\parheading{Privacy Comfort:} Level of privacy comfort is affected by age and by the belief that IoT devices track daily activities. The age groups ``25 - 34'' and ``45 - 54'' had the lowest comfort levels. In addition, the more that users think IoT devices track their daily activities, the lower their level of privacy comfort.

\parheading{Used Blockers:} When examining the correlation between education level and whether one used ad blockers, the Kruskal-Wallis test indicates that there is a statistically significant difference between them. However, Dunn’s test does not reveal statistical significance after conducting pairwise comparisons between education groups themselves. This could be due to the high variability within each education group in ad blocker usage, which makes it difficult to find statistically significant differences between particular pairs. Moreover, it is important to acknowledge the type II errors that might arise from this, in which actual differences exist but are not found to be statistically significant. Lastly, ad blocker usage may be influenced by interaction effects with other variables.

\parheading{Extra Steps:} There is statistical significance in taking steps to block trackers when it pertains to the user's belief in IoT tracking, level of privacy comfort with IoT devices, and previous use of ad/tracker blockers, suggesting a correlation between these variables. Those who believe that IoT devices track their activities would put in the effort to block trackers. Similarly, those who are not comfortable when using IoT devices are more likely to take protective actions and block trackers. Ad blocker users may have stronger feelings toward data privacy, which translates into taking extra steps to block trackers.

\parheading{Would Use IoT Ad/Tracker Blockers, Blocking vs. Usability:} The probability of using IoT ad/tracker blockers and using them despite usability issues are also affected by IoT work experience. Users with IoT work experience will have more technical expertise, giving them more awareness of the level of data collection by IoT devices. This leads them to be willing to install an IoT ad/tracker blocker even if it affects the usability.

\parheading{New Info Learned:} Learning new information has a significant statistical correlation depending on education level and belief that IoT devices track daily activities. Dunn's test shows that there is a significant difference between users with an education level of "Less than high school" and all other groups. The former group learned less new information than the latter. This could be attributed to the information being advanced to an extent where the users were confused and did not understand the content shown in \tool{}.

\parheading{Willingness to Install \tool{}:} There is a notable statistical correlation between the willingness to install \tool{} and the belief that IoT devices track daily activities. Performing Dunn's test shows that users who strongly believe that IoT devices track their daily activities are more likely to install \tool{} than those who do not with $p = 0.058$.

\subsection{Usability of \tool{} (UQ2)}\label{sec:results_vbit_usability}

\subsubsection{\textbf{Acceptable Usability}}
\tool{} achieves acceptable usability for both MR and web apps. Interviewees' SUS scores are computed based on a 100-point scale (Figure \ref{fig:mr_vs_web_SUS}) and evaluated in accordance with Brooke~\cite{sus_brooke} and Bangor~\cite{sus_bangor}. A score of 60 or above is considered acceptable with high marginal. The SUS scores for the interviewees were 77.5 for the MR app and 82.5 for the web app. This places both apps in the acceptable range, with the web app ranking higher than the MR app. This may be due to participants having had more experience with websites in general as opposed to MR apps, and thus the web app may have been initially easier to navigate without assistance or explanation.

All interviewees had positive feedback regarding both apps. Interviewees enjoyed using the MR app because it provides a unique visualization and interaction experience, as I6 explained: \textit{``I really like this in terms of its functionality and this kind of new interface because it’s AR.''} 
Interviewees were also pleased with the level of control that the selective blocking feature provides, which I10 explained: \textit{``That’s really convenient because then you can figure out what you can get away with blocking really easily without completely breaking the functionality of whatever device you’re using. So at least that way you can minimize the amount of trackers that can follow you around, even if you can’t block all of them to maintain functionality. That’s pretty cool.''}

Figure~\ref{app_fig:total_opinions} in Appendix~\ref{app:survey_results} shows the opinions of survey participants on the importance of each app's components. We observed that the non-tracker domains are less important and blocking/unblocking tracker domains was easy for all study participants. In addition, survey participants are more likely to rate the list of trackers in the MR app as more important compared to that of the web app ($p=0.042$). In both apps, the majority of survey participants ($>80\%$) stated that the tracker domains (list of trackers and top tracker domains contacted) are very/extremely important. 

\subsection{Design Insights for \tool{} (UQ3)}\label{sec:results_vbit_design}
The most common suggestions for improvement were related to \emph{transparency} (\eg{} adding more information), \emph{control}, (\eg{} flexible options for blocking), and \emph{customization} (\eg{} custom layouts and settings). These themes are in line with prior work regarding users' expectations for their IoT privacy management, as discussed in Section~\ref{sec:relatedwork}. Survey participants and interviewees generally had similar suggestions. We include the full list of suggestions provided by the survey participants in Figure \ref{app_fig:features_all} in Appendix~\ref{app:survey_results}. Since the interviewees had the opportunity to explore \tool{} in person, we were able to discuss in great detail how they would use \tool{} and their expectations. 

\subsubsection{\textbf{MR App Improvements}}
Many interviewees (11/18) enjoyed the interactive visualization aspect of the MR app, but they found it tiring to continuously hold the phone pointed at the marker. Thus, interviewees suggested having an initial MR aspect of the app but once the Recent Information Panel is clicked, the app should open a fully static page, as I4 explained: \textit{``One thing that I like is that it is AR. One thing that I don’t like about it is also that it’s AR. I like that I can interactively say, `Hey, I want to know information about this' and all it takes is pulling up my camera. What I don’t like is that it stays in that mode.''} I11 suggested deploying the app on a head-mounted display: \textit{``If I have a headset, then maybe that could be a good use.''} 
Other suggestions included improving the user interface and information displayed, such as the visibility of the instructions for clicking on the Recent Information Panel, adding domain-to-organization mapping details, and adding a summary of all devices.
Some interviewees (4/18) disliked the MR markers because they would not want to have markers in their homes near their IoT devices, suggesting a marker-less detection feature.

\subsubsection{\textbf{Web App Improvements}} 
Many interviewees (10/18) suggested improving the domain-to-organization mapping (Figure~\ref{app_fig:devices_alluvial} Appendix~\ref{app:web_app}) to improve readability by adding labels to each column or removing the domains column entirely. Improvements to the outgoing traffic graph were also suggested, including adding an outgoing traffic graph for each separate device and having historical outgoing traffic data. 

\subsubsection{\textbf{System-Wide Improvements}} The following suggested improvements apply to both the MR and web apps.

\parheading{Blocking Feature and Instructions:}
The most commonly suggested improvement was to add a more obvious blocking button (\eg{} check box or radio button) next to the tracker domains and a guide that explains this feature in both apps.

\parheading{Domain Lists:} Some interviewees (6/18) wanted a feature to sort the lists of trackers and non-trackers by different columns in the lists (\eg{} access count and blocked status) instead of by time last contacted. Many interviewees (12/18) also desired multiple-select, block-all, and block-by-organization features, such as I7: \textit{``Let's say if someone doesn't have trust or believe in some company, I think it's better if there's an option to block all the Apple Inc. domains.''}

\parheading{Automatic vs. Manual Blocking:} Many interviewees (8/18) expressed that they would personally like to use manual blocking, whereas some (5/18) preferred automatic blocking. I11 preferred manual blocking to avoid usability issues: \textit{``Otherwise it might break a website and I wouldn't know why.''} In contrast, I4 would rely on the developers' suggestions: \textit{``I’ll let you guys decide. Let the creators of this service decide.''} Automatic blocking was also requested by survey participants.

\parheading{Information About Trackers:} Most interviewees (14/18) wanted more details about tracker domains, including data collected, categorization by purpose and organization, and risk level, which would help them make more informed decisions, as I10 explained: \textit{``If there’s a description of what that domain is generally known for collecting, then that’s also very helpful because it will help me decide which ones are critical and have to be completely blocked and which ones are okay.''} I14 explained that they would not personally need the full domain information: \textit{``I don’t see the need for the whole URL being present. The main domain should be sufficient.''} Survey participants also often requested these tracker details.

\parheading{Usefulness of Non-Trackers:} Some interviewees (4/18) did not find the non-tracker information useful because they were more concerned about trackers, but others were interested out of curiosity. I6 shared that the list of non-trackers provides transparency regarding IoT devices: \textit{``Even though I know it’s non-tracker, by knowing this information, I feel like I’m more in control with those devices.''} Alternatively, I14 expressed that they do not need this information if there is no immediate action to be taken: \textit{``I don’t see the point of having a non-tracker here, this could be an additional feature. It need not be shown because I can’t do anything with this information.''}

\parheading{Notifications:} Almost all interviewees (17/18) expressed interest in notifications (\ie{} email, text, or push notification) or reports (\ie{} daily, weekly, or monthly) of new tracker domains contacted. A few interviewees (3/18) did not want new tracker notifications due to notification overload, such as I14: \textit{``If every time a new tracker is found, I would find it annoying because there’s too many notifications.''}

\parheading{Customization:} Customization was a common request by interviewees, such as color and font options to customize the apps to their liking and make them more user-friendly. I3 emphasized how personalization is important to them: \textit{``I like green, so if it can be green, it will be better! But it’s personal preference.''} Given how interviewees often had varying opinions about the layout, settings, and level of detail desired in the apps, customization was an important aspect. For example, an advanced mode could enable more detailed information if desired whereas the default mode would have fewer details and higher-level summaries with actionable information.

%% file: sections/06-discussion.tex
\section{Conclusion and Future Directions}\label{sec:discussion}

\parheading{Summary.} As the spaces we live and work in become increasingly instrumented with IoT devices, it is important to provide transparency and control over tracking by these devices. In this paper, we present \tool{}, a comprehensive and powerful approach to visualize and control the network traffic generated by IoT devices in real-time.  \tool{} leverages the universal vantage point of network traffic on the edge router, with negligible overhead, and provides user-friendly visualizations and blocking features via both MR and web apps.  Our user study shows that \tool{} can educate users, empower them to make informed privacy decisions, and provide insights for future directions.

\parheading{Deployment and Uses.} 
We envision that \tool{} will be deployed by the administrator of a smart space to offer users transparency and control over tracking by IoT devices in that space.  
In this setting, \tool{}'s components trust each other and work together: the administrator of the smart space owns the devices and the router, which talk to the MR and web apps through a local server to offer the user a comprehensive visualization and options for blocking trackers.
We also envision different sets of users of \tool{}. {\em Lay users} walking into a smart space can appreciate \tool{}'s user-friendly visualizations and blocking features, available as both mobile and web apps, for their device. {\em Administrators of the smart space} can utilize \tool{} for more in-depth analysis and control, such as discovering and blocking previously unknown tracking activities. Researchers can develop additional features (\eg~ IoT-specific block lists) for \tool{}, or other systems based on our design insights and conduct user studies using \tool{} to analyze different research questions regarding users' IoT privacy. {\em Filter list creators} can utilize \tool{} to easily block domains and check for breakage on the corresponding IoT device, thus enabling the creation of more fine-grained and effective IoT blocklists, compared to the ones available today~\cite{Firebog-SmartTV, Firebog-AmazonFireTV}. {\em Regulators and enforcement agencies} can utilize \tool{} to audit the data collection practices and compliance of IoT devices with privacy laws. We envision that adoption in real smart homes and buildings will be relatively easy since (i) \tool{} can be deployed on any Linux-operated access point/router, requiring no extra hardware, and (ii) users can simply download a compatible app for smart homes or smart buildings that run \tool{}. In contrast, tools based on ARP spoofing,  such as \cite{ndss_2023_ar_privacy} and \cite{iot_inspector_2020}, are not allowed on commercial app stores and are published only through the developers' websites.   

Since, our current prototype is for RPi and Pi-hole, a potential beta-testing audience could be open-source communities such as Pi-hole and filter list authors. \tool{} can be released as a standalone package or as an extension to Pi-hole allowing for easier installation. Deploying \tool{} in the wild will also allow us to crowd-source feedback and the labels needed for building robust anti-tracking filter lists. More specifically, \tool{} enables users to test and label which domains can be blocked without breaking functionality for a specific IoT device. This type of ground truth labeled data is very difficult to obtain for IoT (while it is easier for web ad blocking), and is part of why there are no widely accepted filter lists for IoT tracker blocking today. Similar problems are present in the web as mentioned in~\cite{autofr}, where most blocklists are still manually curated.

\parheading{Limitations and Future Work.}
First, we plan to incorporate suggestions from the user study to further {\em improve the design} of \tool{}. Examples of features to consider adding include: network traffic decryption to visualize data collection, per-data type blocking, and developing marker-less detection, possibly accompanied by indoor navigation to find the location of the IoT devices.
Second, \tool{} relies on {\em filter lists} for labeling domains as potential trackers for users' consideration to block. The curation of the filter lists themselves was out of scope in this work. However, \tool{} can be used as a tool to crowd-source labeling of domains as IoT trackers when users select domains to block on our MR and web apps.
Third, and related to the above, 
long-term  {\em in-the-wild deployment}, \eg~ by making \tool{} available to the Pi-hole community, can enable a longitudinal study and guide future improvements. Fourth, in addition to end-users that were the focus of this study, \tool{} can be utilized by {\em experts}, including smart space administrators, filter list authors, or regulators/auditors. Finally, both our survey and interview participants were based in the US, and thus future work should further explore other countries and backgrounds.

%% file: sections/appendix.tex
\newpage

\section{System Design}

In this appendix, we provide more details on \tool{}'s system design and system performance supplementing Section~\ref{sec:design}.
Appendix~\ref{app:mobile_app} and \ref{app:web_app} show extra information regarding the MR and web apps, respectively. Appendix~\ref{app:power_rpi} expands more on the system performance measurements for \tool{} and the mobile phone.

\subsection{MR Mobile App Communication Protocols}\label{app:mobile_app}

\begin{figure}[t]
\centering
\includegraphics[ width=0.65\linewidth]{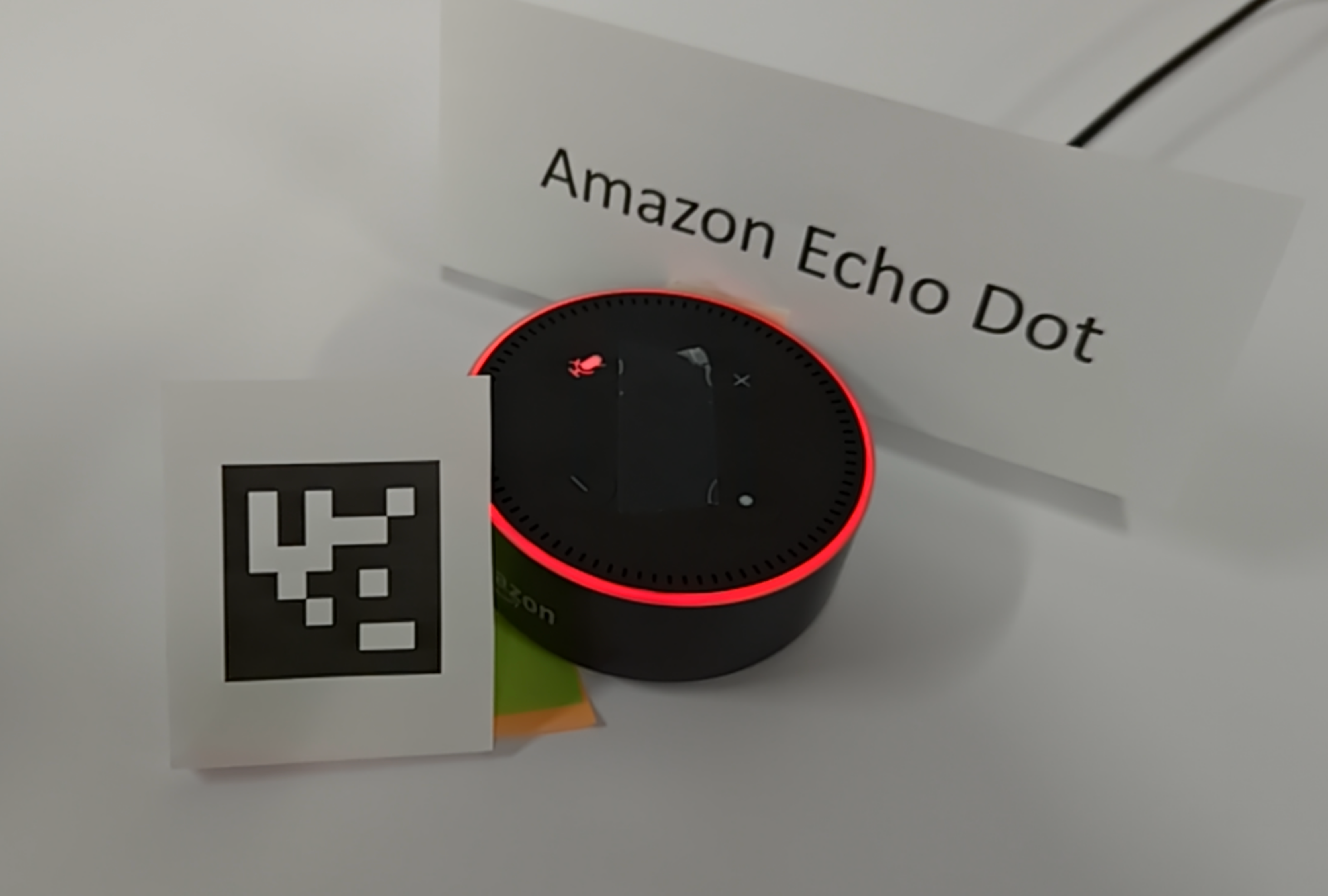}
\caption{ArUco marker ID associated with the Amazon Echo Dot.}
\label{appendix:UI_flow_chart_step1}
\end{figure}

Section~\ref{subsec:mobile_app} discusses \tool{}'s MR mobile app in detail. In this appendix section,
Figure~\ref{appendix:UI_flow_chart_step1} shows the per-device unique ArUco marker that is used to identify each IoT device.

\subsection{Web App Supplemental Material}\label{app:web_app}

\begin{figure*}[t]
\centering
\includegraphics[ width=0.95\linewidth]{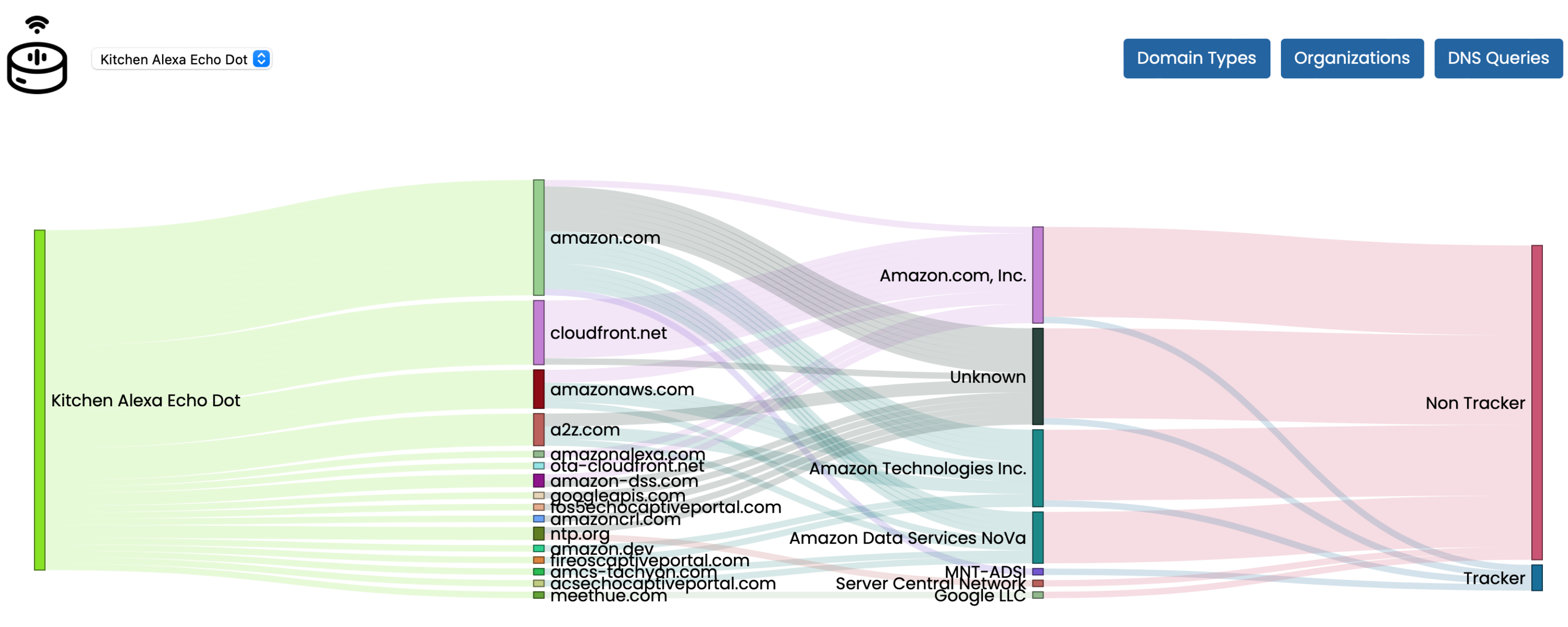}
\caption{Domain-to-organization mapping alluvial graph in devices page. Each of the columns shows: (1) IoT device, (2) SLD domains contacted, (3) organization of each domain, and (4) tracker/non-tracker status.}
\label{app_fig:devices_alluvial}
\end{figure*}

Section~\ref{subsec:web_app} discusses the different components of the web app. In Figure~\ref{app_fig:devices_alluvial}, we show the alluvial diagram, on the devices page, that visualizes the domain-to-organization mapping.

\subsection{Power Consumption}\label{app:power_rpi}
In Section~\ref{subsec:performance} we analyze the system performance. In this section we analyze the power consumption of the RPi.

\parheading{Power Consumption.}
For the Raspberry Pi 4, we used a USB power meter~\cite{usbpowermeter} also with the three scenarios (mentioned in Section~\ref{subsec:performance}). In this case, Pi-hole is the \tool{}'s component that runs on Raspberry Pi. Without \tool{}, with \tool{}, and with \tool{} running and blocking/unblocking traffic, the power consumption is 2.86, 3.49, and 3.85 W, respectively. Thus, Pi-hole reasonably consumes around 22 - 35\% more power at peak activity when blocking and unblocking traffic.

\section{Survey Questionnaire \label{appendix:questionaire}}

\Qitem{ \Qq{Consent}}

\section*{Demographics Information}

\Qitem{ \Qq{What is your age?}
\begin{QlistC}
\item Under 18
\item 18 - 24
\item 25 - 34
\item 35 - 44
\item 45 - 54
\item 55 - 64
\item 65 - 74
\item 75 - 84
\item 85 or older
\end{QlistC}
}

\Qitem{ \Qq{What is your highest education level achieved/pursuing?}
\begin{QlistC}
\item Less than high school
\item High school graduate
\item Some college
\item 2 year degree
\item 4 year degree
\item Professional Degree
\item Master's degree
\item Doctorate
\end{QlistC}
}

\Qitem{ \Qq{What is your gender?}
\begin{QlistC}
\item Male
\item Female
\item Other (please specify): \Qline{2cm}
\item Prefer not to say
\end{QlistC}
}

\Qitem{ \Qq{Have you ever worked in the Internet of Things (IoT), security, and/or privacy fields?}
\begin{QlistC}
\item No
\item Yes, please describe your work: \Qline{2cm}
\end{QlistC}
}

\section*{Pre-Video - IoT Devices}

\Qitem{ \Qq{Which IoT devices do you own? Select all that apply.}
\begin{Qlist}
\item Motion Detector
\item Smart Lock
\item Smartwatch
\item Voice Assistant
\item Smart Thermostat
\item Smart TV
\item Smart Refrigerator
\item Smart Bulb
\item None
\item Others: \Qline{2cm}
\end{Qlist}
}

\Qitem{ \Qq{Explanation of Targeted Advertising.}}

\Qitem{ \Qq{Please answer the following questions regarding IoT devices.}

\begin{enumerate}[label=\arabic*)]
    \item Have you ever experienced targeted advertising with any of the IoT devices you own? \textbf{\textit{Only shown to IoT device owners}}
    
    None at all \QratingC{5} A great deal
    
    \item Do you think IoT devices track your daily activities to send you personalized ads/services?
    
    None at all \QratingC{5} A great deal
\end{enumerate}
}

\Qitem{ \Qq{Please answer the following question.}

\begin{enumerate}[label=\arabic*)]
    \item Rate your level of privacy comfort when using IoT devices.
    
    Extremely uncomfortable \QratingC{5} Extremely comfortable
\end{enumerate}
}

\section*{Pre-Video - Ads and Blockers}

\Qitem{ \Qq{Explanation of ad blocking and web tracking.}}

\Qitem{ \Qq{Have you ever used any ads and trackers blockers?}
\begin{QlistC}
\item No
\item Yes, please mention where/what platforms: \Qline{2cm}
\end{QlistC}
}

\Qitem{ \Qq{Please answer the following questions about ads and tracker blockers (for example ads and trackers blocker plugins for browsers).}

\textbf{Scale for all sub-questions:} Extremely unlikely \QratingC{5} Extremely likely

\begin{enumerate}[label=\arabic*)]
    \item Would you take extra steps (such as installing extra software plugin/hardware devices) to detect/block possible trackers?
    \item Would you use a system that could block tracking services on your IoT devices?
    \item Would you use a system  that could block tracking services if it affected the usability of the device? For example, if it slowed down the response time of the device or if it resulted in breakage of some of the device functionality.
\end{enumerate}
}

\section*{MR App - Video Questions}

\Qitem{ \Qq{Explanation of domain.}}

\Qitem{ \Qq{System Overview Figure}}

\Qitem{ \Qq{VBIT Video: MR Application}}

\Qitem{ \Qq{Four labeled figures from the MR application.}}

\Qitem{ \Qq{MR Application - Please answer the following questions based on the importance of the information shown in the screenshots above.}

\textbf{Scale for all sub-questions:} Not at all important \QratingC{5} Extremely important

\begin{enumerate}[label=\arabic*)]
    \item Recent information Panel
    \item Domain Information Shown
    \item List of non-trackers
    \item List of trackers
\end{enumerate}
}

\Qitem{ \Qq{GIF of blocking and unblocking a tracker domain in the MR app.}}

\Qitem{ \Qq{How easy is it to block/unblock tracker domains?}

Extremely difficult \QratingC{5} Extremely easy
}

\Qitem{ \Qq{Attention check question.}}

\Qitem{ \Qq{Are there any features, functionalities, or informational descriptions that you would add to the MR application to help you understand the MR application better and learn how to use it?}

Answer: \Qline{2cm}
}

\section*{Web App - Video Questions}

\Qitem{ \Qq{Explanation of domain.}}

\Qitem{ \Qq{VBIT Video: Web Application}}

\Qitem{ \Qq{Five labeled figures from the Web application - Dashboard Page.}}

\Qitem{ \Qq{Web Application - Dashboard Page - Please answer the following questions based on the importance of the information shown in the screenshots above.}

\textbf{Scale for all sub-questions:} Not at all important \QratingC{5} Extremely important

\begin{enumerate}[label=\arabic*)]
    \item Outgoing traffic graph
    \item Top non-tracker domains contacted
    \item Top tracker domains contacted
    \item Domains contacted per device pie chart
    \item Domain to organization chart
\end{enumerate}
}

\Qitem{ \Qq{Six labeled figures from the Web application - Device Page.}}

\Qitem{ \Qq{Web Application - Device Page - Please answer the following questions based on the importance of the information shown in the screenshots above.}

\textbf{Scale for all sub-questions:} Not at all important \QratingC{5} Extremely important

\begin{enumerate}[label=\arabic*)]
    \item Domain types pie chart
    \item Domain to organization chart
    \item DNS queries chart
    \item Recent domains contacted table
    \item List of non-trackers
    \item List of trackers
\end{enumerate}
}

\Qitem{ \Qq{GIF of blocking and unblocking a tracker domain in the web app.}}

\Qitem{ \Qq{How easy is it to block/unblock tracker domains?}

Extremely difficult \QratingC{5} Extremely easy
}

\Qitem{ \Qq{Are there any features, functionalities, or informational descriptions that you would add to the Web application to help you understand the Web application better and learn how to use it?}

Answer: \Qline{2cm}
}

\section*{Post Video - Questions}

\Qitem{ \Qq{Please describe what new information about IoT devices you learned from our videos.}

Answer: \Qline{2cm}
}

\Qitem{ \Qq{Would you install such a system, VBIT, at your house to visualize and/or block IoT traffic?}

Extremely unlikely \QratingC{5} Extremely likely
}

\Qitem{ \Qq{After having seen the information displayed by our system, VBIT, please answer the following question again.}

Rate your level of privacy comfort when using IoT devices.

Extremely uncomfortable \QratingC{5} Extremely comfortable
}

\section{Interview SUS Questionnaire \label{appendix:interview_sus}}

\textbf{System Usability Scale:} Interviewees were asked to fill out this SUS questionnaire for the MR and web apps separately.\label{appendix:sus_questions}

\textbf{Scale for all sub-questions:} Strongly disagree \QratingC{5} Strongly agree

\begin{enumerate}[label=\arabic*)]
    \item I think that I would like to use this system frequently.
    \item I found the system unnecessarily complex.
    \item I thought the system was easy to use.
    \item I think that I would need the support of a technical person to be able to use this system.
    \item I found the various functions in this system were well integrated.
    \item I thought there was too much inconsistency in this system.
    \item I would imagine that most people would learn to use this system very quickly.
    \item I found the system very cumbersome to use.
    \item I felt very confident using the system.
    \item I needed to learn a lot of things before I could get going with this system.
\end{enumerate}

\section{User Study Supplemental Material}
In this appendix, we provide more detailed data and results for the user study (Section \ref{sec:user-study}). Appendix~\ref{app:demographics} shows the full demographic data of the survey and the interview participants. Appendix~\ref{app:survey_results} provides more details on the results.

\subsection{Demographics Results}\label{app:demographics}

\begin{table}[!htbp]
    \footnotesize
    \centering
    \begin{minipage}[t]{0.49\textwidth}
    \centering
    \caption{Survey participants' demographic information.}
    \begin{tabular}{m{2em}p{15em} r r} \Xhline{2\arrayrulewidth}
         & &  \multicolumn{2}{p{8em}}{\textbf{\makecell{Participants \\ (N=200)}}} \\ \Xhline{2\arrayrulewidth}

         & & n & \% \\ \cline{3-4}
        \multirow{4}{*}{{\rotatebox[origin=c]{90}{\textbf{Gender}}}} & Male & 95 & 47.5 \\
          & Female & 98 & 49.0 \\
          & Non-binary & 4 & 2.0 \\
          & Prefer not to answer & 3 & 1.5 \\ \hline
        \multirow{5}{*}{{\rotatebox[origin=c]{90}{\textbf{Age}}}} & 18-24 years old & 27 & 13.5 \\
          & 25-34 years old & 74 & 37.0 \\
          & 35-44 years old & 57 & 28.5 \\
          & 45-54 years old & 22 & 11.0 \\
          & 55-64 years old & 13 & 6.5 \\
          & 65 years old or older & 7 & 3.5 \\ \hline
        \multirow{9}{*}{{\rotatebox[origin=c]{90}{\textbf{Education}}}} & Less than high school & 3 & 1.5 \\
          & High school graduate & 21 & 10.5 \\
          & Some college & 35 & 17.5 \\
          & 2-year college degree & 29 & 14.5 \\
          & 4-year college degree & 76 & 38.0 \\
          & Master's & 28 & 14.0\\
          & Doctorate/Professional & 8 & 4.0 \\ 
          \Xhline{2\arrayrulewidth}
    \end{tabular}
    \label{app_tab:demographic_survey}
    \end{minipage}
    \begin{minipage}[t]{0.49\textwidth}
    \centering
    \caption{Interview participants' demographic information.}
    \begin{tabular}{m{2em}p{15em} r r} \Xhline{2\arrayrulewidth}
         & &  \multicolumn{2}{p{8em}}{\textbf{\makecell{Participants \\ (N=18)}}} \\ \Xhline{2\arrayrulewidth}

         & & n & \% \\ \cline{3-4}
        \multirow{3}{*}{{\rotatebox[origin=c]{90}{\textbf{Gender}}}} & Male & 10 & 55.6 \\
          & Female & 7 & 38.9 \\
          & Prefer not to answer & 1 & 5.6 \\ \hline
        \multirow{2}{*}{{\rotatebox[origin=c]{90}{\textbf{Age}}}} & 18-24 years old & 3 & 16.7 \\
          & 25-34 years old & 15 & 83.3 \\ \hline
        \multirow{3}{*}{{\rotatebox[origin=c]{90}{\textbf{Educ.}}}} & 4-year college degree & 4 & 22.2 \\
          & Master's & 7 & 38.9\\
          & Doctorate & 7 & 38.9 \\ 
          \Xhline{2\arrayrulewidth}
    \end{tabular}
    \label{app_tab:demographic_interview}
    \end{minipage}
\end{table}

\subsection{Survey Results}\label{app:survey_results}

Please recall that Section~\ref{sec:results} analyzed the survey and interview results. This appendix expands more on the results.

\begin{figure}[t]
\renewcommand{\thesubfigure}{\alph{row}.}
\setcounter{row}{1}
\centering
\begin{subfigure}{.40\textwidth}
    \centering
    \includegraphics[width=1\linewidth]{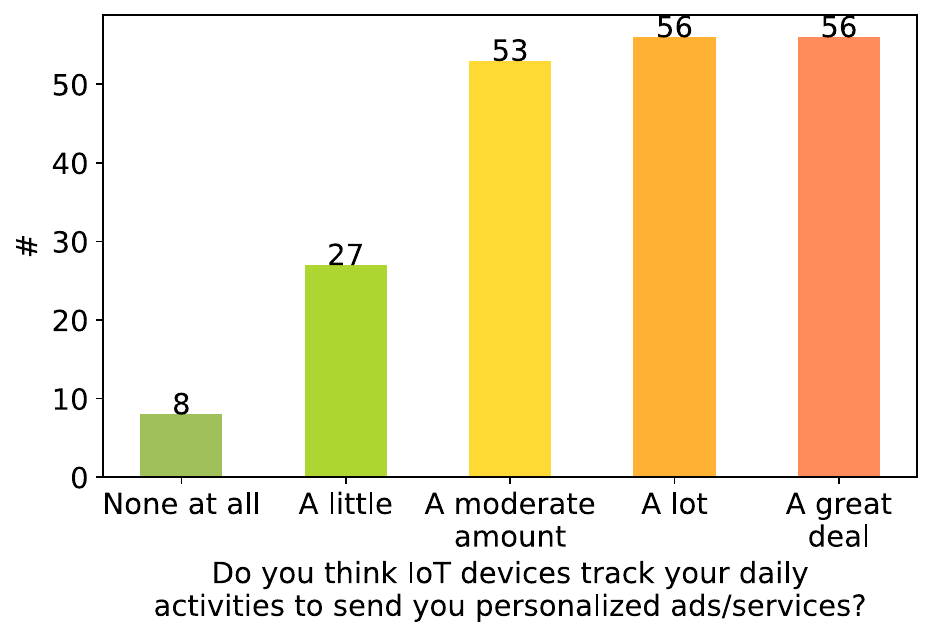}
    \caption{Survey participants' beliefs regarding IoT tracking.}
    \label{fig:belief_tracking}
\end{subfigure}
\stepcounter{row}
\begin{subfigure}{.45\textwidth}
    \centering
    \includegraphics[width=1\linewidth]{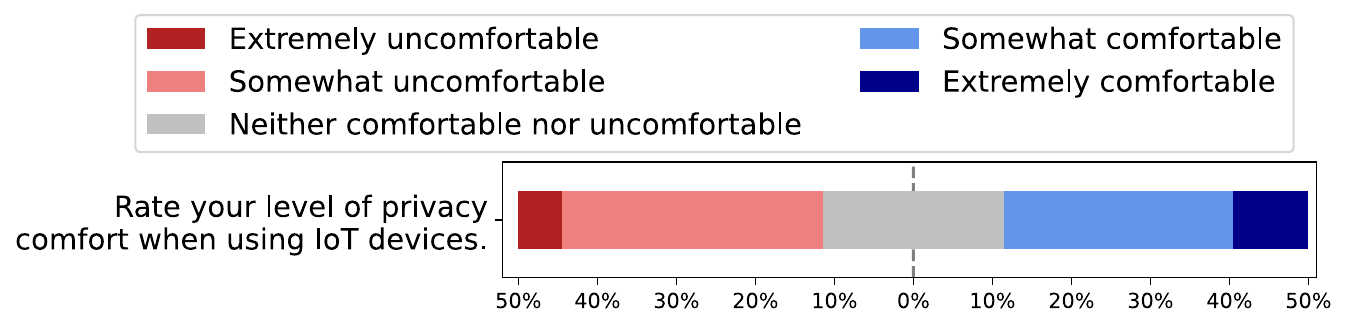}
    \caption{Survey participants' level of privacy comfort.}
    \label{fig:comfort_level_pre}
\end{subfigure}
\caption{Survey participants' perceptions regarding IoT tracking and privacy comfort.}
\end{figure}

\parheading{\textbf{Participants' Privacy Concerns.}}
Participants had varying levels of concern regarding ads and tracking in IoT devices before the \tool{} demo. Many survey participants (63.59\%) experienced at least a moderate amount of targeted ads on owned IoT devices. Most survey participants (82.5\%) believed that IoT devices track daily activities at least a moderate amount to send personalized ads (Figure \ref{fig:belief_tracking}).
Figure~\ref{fig:comfort_level_pre} also shows that about 50\% felt uncomfortable when using IoT devices.

Many interviewees (7/18) disliked that IoTs track users' daily activities, particularly IoTs that collect sensitive data (e.g., audio/visual data from microphones and cameras), as I10 explained: \textit{``It’s creepy. So that’s why I don’t use any of them. Especially if it has a microphone or camera on it, I’m not using it.” } However, a few interviewees (3/18) expressed discomfort with IoT tracking but continue to use their IoT devices because of the utility they provide. I1 explained that while it seems voice assistants are always listening, it does not concern them enough to change their IoT usage: \textit{``If I just say, `Alexa, do this,' then it automatically updates or reacts. So by nature, it feels like they’re always listening for those keywords. So that feels more invasive in that matter. But I still use them, so I guess at the end of the day, it doesn’t really affect me too much.''}

On the other hand, many interviewees (8/18) did not mind IoT tracking and targeted ads because the ads may be useful to them or they feel the tracking is unavoidable, as long as they still have some control. I7 is willing to accept IoT tracking if their data is not shared: \textit{``It’s okay as long as it’s not being sold to some third-party sites which try to exploit this. As long as it’s being used to give useful ads to me, I guess it’s okay.''}
Additionally, I4 explained the inevitability of IoT tracking: \textit{``For people with something to hide, maybe it matters, but most companies or the really big companies, if they want my information, they can get it. I feel like preventing them from tracking it through IoT devices won’t stop them.''}

\parheading{IoT Devices Usage.} Most of the survey participants own a Smart TV (77.0\%) and a Voice Assistant (52.5\%), and only 8.0\% of the survey participants do not own an IoT device. Table~\ref{app_tab:iot_devices_owned} shows the statistics on what IoT devices our survey participants own.

\begin{table}[t]
    \footnotesize
    \centering
    \caption{Types of IoT devices owned by survey participants.\newline}
        \begin{tabular}{p{0.4\linewidth}p{0.2\linewidth}p{0.2\linewidth}}
        \Xhline{2\arrayrulewidth}
        \textbf{IoT Device} & \textbf{n} & \textbf{\%}\\
        \Xhline{2\arrayrulewidth}
        Smart TV                  &  154 & 77.0\\
        Voice Assistant           &  105 & 52.5\\
        Smartwatch                &   92 & 46.0\\
        Smart Bulb                &   50 & 25.0\\
        Smart Thermostat          &   44 & 22.0\\
        Motion Detector           &   41 & 20.5\\
        Smart Lock                &   23 & 11.5\\
        Smart Refrigerator        &   11 & 5.5\\
        None                      &   16 & 8.0\\
        Others                    &   12 & 6.0\\
        \Xhline{2\arrayrulewidth}
        \end{tabular}
        \label{app_tab:iot_devices_owned}
\end{table}

\begin{figure}[t]
\centering
\includegraphics[width=0.75\linewidth]{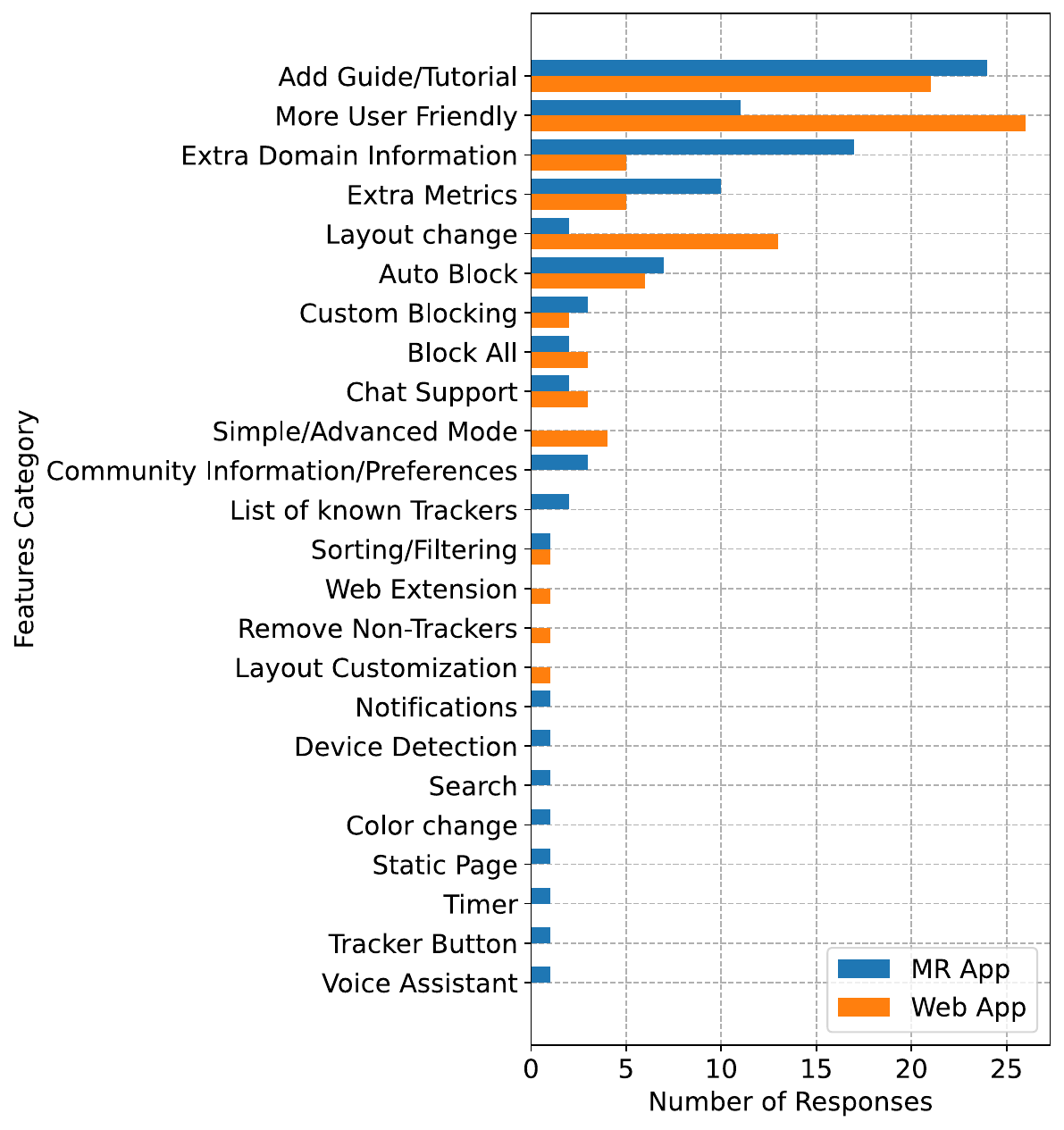}
\caption{MR vs. web app features requested by survey participants.}
\label{app_fig:features_all}
\end{figure}

\begin{figure}[t]
\centering
\includegraphics[width=0.75\linewidth]{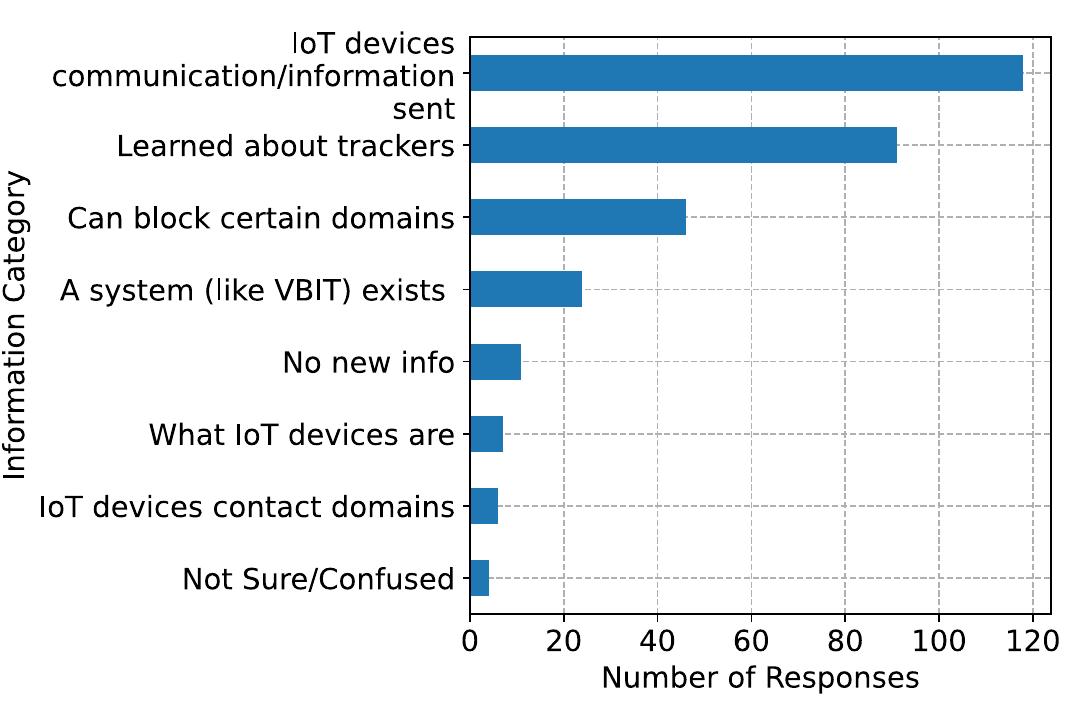}
\caption{New information learned from the \tool{} system for survey participants.}
\label{app_fig:new_info_learned}
\end{figure}

\begin{figure}[t]
\centering
\includegraphics[width=0.75\linewidth]{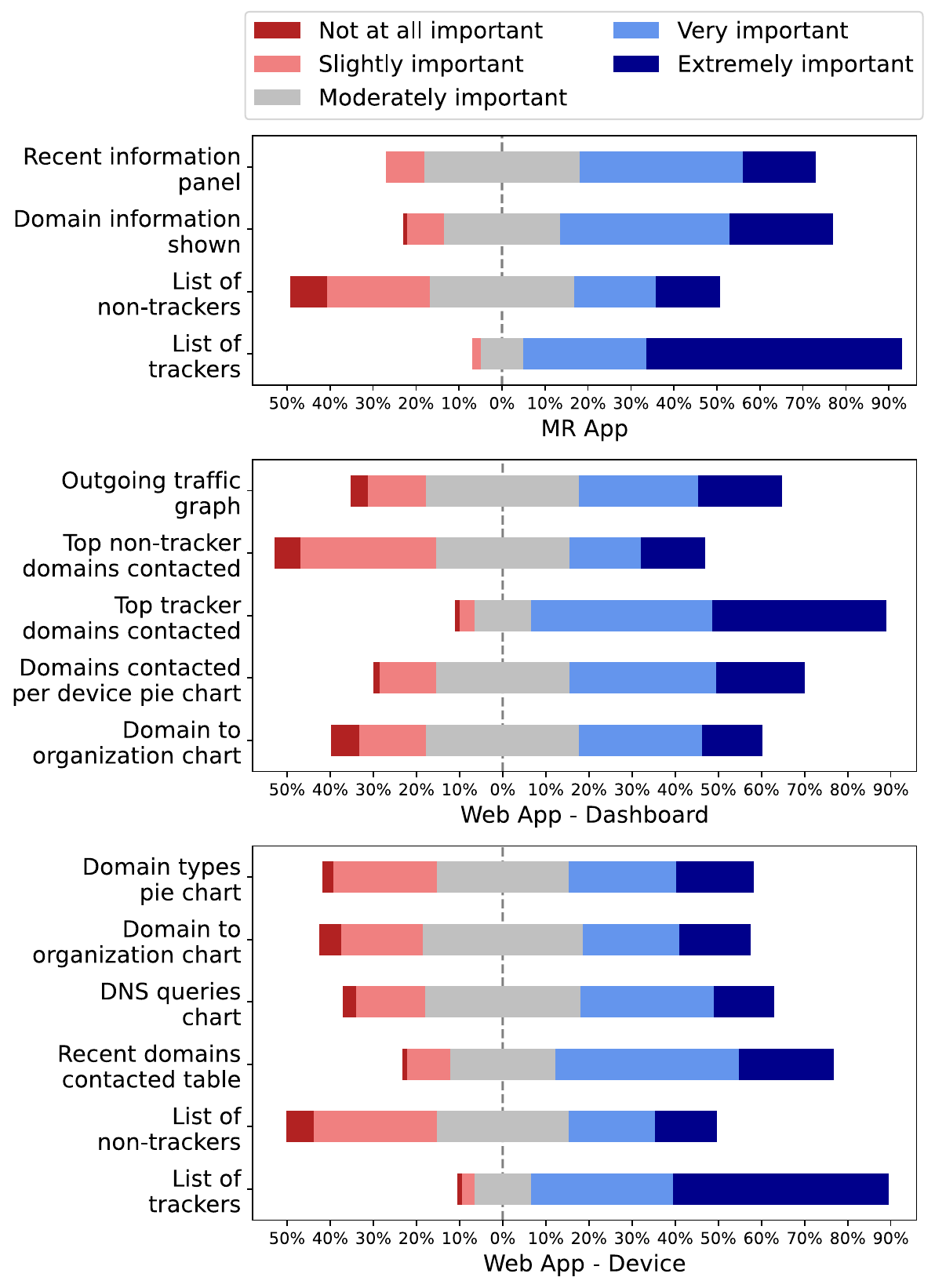}
\caption{Importance of each component of the MR and web apps for survey participants.}
\label{app_fig:total_opinions}
\end{figure}

\parheading{Survey Results.} 
Figure~\ref{app_fig:features_all} highlights the category of features requested by survey participants to be added to \tool{}.
Figure~\ref{app_fig:new_info_learned} shows the new information learned by the survey participants. 
Figure~\ref{app_fig:total_opinions} shows the full list of opinions of the survey participants regarding the components of the MR and web apps.

